\documentclass[11pt,draftclsnofoot,onecolumn]{IEEEtran}

\usepackage{amssymb}
\usepackage{amsmath, amsthm}
\usepackage[dvips]{graphicx}
\usepackage[font=footnotesize]{subfig}
\usepackage{multirow}

\newtheorem{proposition}{Proposition}

\title{A Game Theoretic Analysis of Incentives\\in Content Production and Sharing\\over Peer-to-Peer Networks}
\author{Jaeok Park and Mihaela van der Schaar\thanks{The authors are with Electrical Engineering Department, University of
California, Los Angeles (UCLA), 420 Westwood Plaza,
Los Angeles, CA 90095-1594, USA. e-mail:
\{jaeok, mihaela\}@ee.ucla.edu.}
}
\date{}

\begin{document}

\maketitle


\begin{abstract}
User-generated content can be distributed at a low cost using peer-to-peer
(P2P) networks, but the free-rider problem hinders the utilization
of P2P networks. In order to achieve an efficient use of P2P networks,
we investigate fundamental issues on incentives in content production and sharing using
game theory.
We build a basic model to analyze non-cooperative outcomes without
an incentive scheme and then use different game formulations derived
from the basic model to examine five incentive schemes: cooperative,
payment, repeated interaction, intervention, and enforced full sharing.
The results of this paper show that
1) cooperative peers share all produced content
while non-cooperative peers do not share at all without an incentive
scheme; 2) a cooperative scheme allows peers to consume more content
than non-cooperative outcomes do; 3) a cooperative outcome can be
achieved among non-cooperative peers by introducing an incentive scheme based on
payment, repeated interaction, or intervention; and 4) enforced full sharing
has ambiguous welfare effects on peers.
In addition to describing the solutions of different formulations, we
discuss enforcement and informational requirements to
implement each solution, aiming to offer a
guideline for protocol designers when
designing incentive schemes for P2P networks.
\end{abstract}

\begin{IEEEkeywords}
Game theory, incentives, network economics, peer-to-peer networks, pricing schemes.
\end{IEEEkeywords}

\section{Introduction}

Recent developments in technology have significantly reduced the cost of producing and
distributing content in various forms such as images, sounds, videos, and text.
Once produced only by companies with a large capital, content can
now be produced by end-users.
In today's Internet-based social communities, peer-to-peer (P2P) networks
offer a cost effective and easily deployable framework for sharing
user-generated content \cite{hgpark}. While P2P networks have
many advantages such as scalability, resilience, and
effectiveness in coping with dynamics and heterogeneity \cite{jliu},
they are vulnerable to intrinsic incentive problems in
that the transfer of content incurs costs to uploaders as well as
to downloaders but benefits only downloaders.
Since the social cost of transfer (the sum of upload and download costs)
exceeds the private cost of transfer (download costs),
peers tend to download excessibly
as in the tragedy of the commons problem. On the other hand, since
upload incurs costs to uploaders without giving them direct benefit,
peers tend to upload too little.
The incentive problem stating that peers desire to benefit from P2P networks while
not contributing to them is referred to as the free-rider (or freeloader) problem.

Various incentive schemes to mitigate the free-rider problem have been
proposed and analyzed in the literature. Cooperative schemes (e.g., \cite{garb}, \cite{wang})
utilize helpers that download files on behalf of a peer in the same collaborative
group. Helpers can improve the download performance of P2P networks
by sharing their spare upload capacities. However, forming and sustaining
collaborative groups in a distributed system poses a main challenge to cooperative schemes.
Payment schemes (e.g., \cite{vishn}, \cite{siri}) use virtual currency
or micropayment to reward upload and charge download. Payment schemes have
a solid theoretical foundation as they are based on economic models.
However, they are regarded impractical because they require an accounting
infrastructure to track the transactions of peers \cite{bura}.
Differential service schemes (e.g., \cite{kamvar}, \cite{xiong}) allow peers to make upload decisions
based on the rating of a peer that requests content from them.
Since a peer with a good reputation is treated preferentially by other
peers, differential service schemes provide incentives for peers to contribute
in order to build and maintain a good reputation. However, differential service schemes
require large communication overheads to determine and announce the ratings
of peers. The rating of a peer is determined by its past actions, which are
observed by different peers, and peers need to know the rating
of every other peer that they interact with.

Game theory \cite{fudenberg} offers a useful framework to model multi-user
interaction and has been applied to analyze the behavior of peers in P2P networks.
Incentive schemes such as payment schemes and differential service schemes
have been investigated using non-cooperative game theory.
Payment schemes can be easily incorporated in static game models
as in \cite{golle}, while differential service schemes have been studied
in the context of different game models.
\cite{bura} uses a static game model to analyze the steady-state outcome
of learning dynamics under a differential service scheme.
\cite{feldman} simulates an evolutionary game model to examine the performance
of a differential service scheme based on peer reciprocation.
\cite{blanc} uses a repeated game model to construct a differential service
scheme based on the idea of social norms \cite{kandori}.
\cite{ma} and \cite{loginova} apply the mechanism design
approach to build optimal incentive-compatible differential service
schemes. \cite{wlin} uses both repeated game and mechanism design approaches
to propose cheat-proof and attack-resistant differential service
schemes. Cooperative game theory has also been used to investigate
coalition formation among peers \cite{yeung}, \cite{hgpark2}.

In this paper, we investigate fundamental issues on incentives
in content production and sharing over P2P networks using game theory.
Unlike existing game-theoretic works on P2P networks, which focus on
a particular game model to construct incentive schemes, we
build a basic model and use it as a unified framework based on which
different incentive schemes are examined applying various
game theoretic models. Specifically, we analyze the basic model
as a non-cooperative game and examine five incentive schemes --- cooperative,
payment, repeated interaction, intervention, and enforced full sharing ---
using different game formulations derived from the basic model,
as summarized in Table I.
Hence, instead of arguing for a particular incentive scheme and a modeling
approach, we show that alternative incentive schemes can provide
incentives for sharing in P2P networks from a neutral perspective.
As can be seen from Table I, different incentive schemes and the
corresponding game models have different requirements for implementation.
Since the characteristics of P2P networks vary depending on their
architecture, the effectiveness of an incentive scheme will
depend on the network environment. Thus, our analysis in this paper
can serve as a guideline for a protocol designer when modeling,
comparing, and selecting incentive schemes.

\begin{table}
\caption{Comparisons of the approaches discussed in the paper. $\mathbf{v} \triangleq (v_1, \ldots, v_N)$
represents the utility functions of peers.
$\mathbf{y}^* \triangleq (y_1^*, \ldots, y_N^*)$
represents the Pareto efficient sharing levels of peers desired by the protocol designer (PD).
$t^*$ and $q^*$ represent optimal payment and intervention schemes, respectively,
that implements $\mathbf{y}^*$. Rationality of peers are assumed throughout.
}
\centering
\begin{tabular}{|c||c|c|c|c|}
\hline
Section & Approach & Requirements (enforcement and information) & Pareto efficiency & Prop. \\
\hline \hline
III & Non-cooperative & There are no requirements for the PD and peers. & Inefficient & 1 \\
\hline
& & The PD knows $\mathbf{v}$ to compute $\mathbf{y}^*$. & & \\[-1ex]
IV & Cooperative & The PD enforces $\mathbf{y}^*$. & Efficient & 4\\[-1ex]
& & Peer $i$ knows $\mathbf{y}^*$. & &\\
\hline
& & The PD knows $\mathbf{v}$ to compute $t^*$.  & & \\[-1ex]
V & Payment & The PD enforces $t^*$. & Efficient & 6\\[-1ex]
& & Peer $i$ knows $t^*$. & &\\
\hline
& Differential service & The PD knows $\mathbf{v}$ to compute $\mathbf{y}^*$. & & \\[-1ex]
VI.A & based on & Peer $i$ knows $\mathbf{y}^*$ and the punishment rule, & Efficient & 7\\[-1ex]
& repeated interaction & and maintains a history of past observations. & &\\
\hline
& Differential service & The PD knows $\mathbf{v}$ to compute $q^*$.  & & \\[-1ex]
VI.B & based on & The PD enforces $q^*$. & Efficient & 8\\[-1ex]
& intervention & Peer $i$ knows $q^*$. & &\\
\hline
& Enforced & The PD enforces full sharing (not sharing levels). & & \\[-1ex]
\raisebox{1.5ex}{VII} & full sharing & Peer $i$ knows $\mathbf{v}$ for endogenous network formation. & \raisebox{1.5ex}{Inefficient} & \raisebox{1.5ex}{9, 11} \\
\hline
\end{tabular}
\end{table}

Another distinctive feature of our framework is that we allow peers to make production decisions
whereas most of existing works assume that peers are endowed
with a certain amount of content (see, for example, \cite{golle}, \cite{loginova}).
When produced content and downloaded content
are substitutable in consumption, the amount of content a peer produces is affected by the
amount of content available in a P2P network.
By endogenizing the amount of content that peers produce, we can
capture the strategic link between producing and downloading content.
In addition, we consider scenarios where the sharing decisions of peers
can be enforced while production and download decisions are made in
a non-cooperative manner. These scenarios can be formulated as games with
partial cooperation in which the strategies of players can be enforced
only in some stages. The concepts developed in the discussion of enforced sharing
decisions can be applied to other scenarios that can be modeled
as multi-stage games.

The rest of this paper is organized as follows. In Section II, we formulate
the basic model that describes a scenario of content production and sharing.
In Section III, we analyze the basic model as a non-cooperative game
and identify the free-rider problem. In Section IV, we investigate
cooperative schemes by deriving a
coalitional game based on the basic model. In Section V,
we augment the basic model with a payment scheme to achieve
cooperative outcomes among non-cooperative peers. In Section VI,
we study differential service schemes applying
repeated game and intervention approaches to the basic model. In Section VII,
we analyze a partially cooperative scenario where full sharing
is enforced. In Section VIII, we provide numerical illustration. In Section
IX, we conclude and discuss future directions. Proofs of propositions
are provided either following propositions or in Appendix B.

\section{Model}

We consider a completely connected P2P network of $N$ peers as in \cite{bura}, \cite{golle}.
Peers produce content (e.g., photos, videos, news, and customer reviews) and
use the P2P network to distribute produced content.
Following \cite{jpark}, we model the content production and sharing scenario
as a sequential game consisting of three stages, which is called
the content production and sharing (CPS) game.
\begin{itemize}
\item \emph{Stage One (Production):} Each peer determines its level of
production.\footnote{We use the term production in a broad sense
to mean any method of obtaining content other than
download in the P2P network.}
$x_i \in \mathbb{R}_+$
represents the amount of content produced by peer $i$ and is known
only to peer $i$.

\item \emph{Stage Two (Sharing):} Each peer specifies its level of sharing.
$y_i \in [0,x_i]$ represents the amount of content that peer $i$ makes
available to other peers. $(y_1, \ldots, y_N)$ is known to all peers at the end of stage two.

\item \emph{Stage Three (Transfer):} Each peer determines the amounts of content
that it downloads from other peers. Peer $i$ serves all the requests it receives
from any other peer up to $y_i$. $z_{ij} \in [0,y_j]$ represents the
amount of content that peer $i$ downloads from peer $j \neq i$, or
equivalently peer $j$ uploads to peer $i$.
\end{itemize}

Let $\mathcal{N} \triangleq \{1, \ldots, N \}$ be the set of peers in the P2P network.
For notations, we define $\mathbf{x} \triangleq (x_1, \ldots, x_N)$, $\mathbf{y}
\triangleq (y_1, \ldots, y_N)$, and $\mathbf{Z} \triangleq [z_{ij}]_{i,j \in \mathcal{N}}$,
an $N$-by-$N$ matrix whose $(i,j)$-entry is given by $z_{ij}$, where we set
$z_{ii} = 0$ for all $i \in \mathcal{N}$. The download profile of
peer $i$ is given by the $i$th row of $\mathbf{Z}$, denoted by
$\mathbf{z}_i \triangleq (z_{i1}, \ldots, z_{iN})$. Similarly, the
upload profile of peer $i$ is given by the $i$th column of
$\mathbf{Z}$, denoted by $\mathbf{z}^i \triangleq (z_{1i}, \ldots,
z_{Ni})$. Given the elements of $\mathbf{Z}$, we can compute the
download volume of peer $i$ by $d_i(\mathbf{z}_i) \triangleq \sum_{j
= 1}^{N} z_{ij}$ and its upload volume by $u_i(\mathbf{z}^i)
\triangleq \sum_{j = 1}^{N} z_{ji}$. For notational convenience, we
suppress the dependence of $d_i$ and $u_i$ on $\mathbf{Z}$ and write
$d_i$ and $u_i$ instead of $d_i(\mathbf{z}_i)$ and
$u_i(\mathbf{z}^i)$, respectively. Also, we define $w(\mathbf{Z})$
to be the total transfer volume of the P2P network given $\mathbf{Z}$, i.e.,
$w(\mathbf{Z}) \triangleq \sum_{i=1}^N \sum_{j=1}^N z_{ij} =
\sum_{i=1}^N d_i = \sum_{i=1}^N u_i$, which can be considered as a measure of the
utilization of the P2P network.

We assume that peers produce nonidentical content of homogeneous quality, which
allows us to focus on the quantity of content.
The total amount of content that peer $i$ has at the end of the CPS game, which we call
the consumption of peer $i$, is given by
the sum of the amounts it produces and downloads, $x_i +
d_i$.
The utility of peer $i$ is given by the benefit of
consumption minus the costs of production, download, and
upload:
\begin{align*}
v_i(\mathbf{x}, \mathbf{y}, \mathbf{Z}) = f(x_i + d_i) - \kappa x_i -
\delta d_i - \sigma u_i.
\end{align*}
We analyze the case of homogeneous peers in that $f$, $\kappa$, $\delta$,
and $\sigma$ are the same for all peers.\footnote{We consider homogeneous peers
for analytic tractability. The concepts in this paper can be extended to the case with heterogeneous
peers in a straightforward manner.}
The benefit of consumption is measured by a concave function
$f:\mathbb{R}_+ \rightarrow \mathbb{R}_+$ as in \cite{goyal}. We assume
that $f$ is twice continuously differentiable and satisfies $f(0) =
0$, $f' > 0$ and $f'' < 0$ on $\mathbb{R}_{++}$. We also assume that $f'(0)$ is finite,%
\footnote{We use $f'(0)$ to represent the right derivative of $f$ at 0.}
$f'(0) > \kappa$, and $\lim_{x \rightarrow \infty} f'(x) = 0$ so that for
every $\alpha \in (0,f'(0)]$ there exists a unique $\hat{x}_{\alpha} \geq 0$
that satisfies $f'(\hat{x}_{\alpha}) = \alpha$.
We use linear cost functions as widely adopted in the literature
(see, for example, \cite{bura}, \cite{vdschaar}).\footnote{The linearity of cost functions is assumed for analytic
convenience as it allows us to obtain closed-form expressions for production levels at various solution concepts. The results of this
paper can be easily extended to the case of a general convex cost function of production, as discussed in Sections
III, IV, and V.} The cost of
producing the amount of content $x_i$ is given by $\kappa x_i$, where $\kappa > 0$ is the marginal cost
of production. Download and
upload create costs in terms of bandwidth usage, and transferring the amount of content $z_{ij}$ from peer $i$
to peer $j$ induces a cost of $\delta
z_{ij}$ to peer $i$ (the downloader) and $\sigma
z_{ij}$ to peer $j$ (the uploader), where $\delta > 0$ and $\sigma > 0$ are the marginal costs
of download and upload, respectively. The P2P network has a
positive social value only if obtaining a unit of content through the
P2P network costs less to peers than producing it privately. Hence,
we assume that $\kappa > \delta + \sigma$ to ensure that the P2P network
is socially valuable.

We illustrate the considered scenario with an example of financial
data. Suppose that peers need financial data (e.g., earnings of companies,
gross domestic products, and interest rates) in order to make forecasts
based on which they make investment decisions (e.g., trade stocks
and bonds). To obtain financial data, peers can either collect data by
themselves or download data shared by other peers. A peer can make a more
informed decision with a larger amount of data. Hence, benefits that peers receive from data
can be represented by an increasing benefit function. The benefit function is
concave when the marginal returns of information are
diminishing in the total amount. Alternatively, we can obtain a concave benefit function
by assuming that there is possible duplication in collected data,
peers cannot identify the contents of data before downloading them, and the benefit is
proportional to the amount of distinct data. Appendix A presents a formal proof of
this statement.

\section{Non-Cooperative Analysis}

We first study the non-cooperative outcome of the CPS game, without any incentive
or enforcement device.
Non-cooperative peers choose their strategies
to maximize their own utilities given others' strategies. Thus,
peers' strategies should be self-enforcing at non-cooperative equilibrium in that no peer
can gain by choosing a different strategy unilaterally.
A strategy for peer $i$ in the CPS game is its complete contingent plan over the three stages and is denoted by
$(x_i, y_i(x_i), \mathbf{z}_i(x_i, \mathbf{y}))$. A stage-one strategy for peer $i$ is represented by $x_i \in \mathbb{R}_+$,
a stage-two strategy by a function $y_i:\mathbb{R}_+ \rightarrow \mathbb{R}_+$ such that $y_i(x_i) \leq x_i$ for all $x_i \in \mathbb{R}_+$,
and a stage-three strategy by a function $\mathbf{z}_i:\mathcal{I}_3 \rightarrow \mathbb{R}_+^N$
such that $z_{ij}(x_i, \mathbf{y}) \leq y_j$ for all $j \neq i$ and $z_{ij}(x_i, \mathbf{y}) = 0$ for $j = i$, where
$\mathcal{I}_3 \triangleq \{(x_i, \mathbf{y})|x_i \in \mathbb{R}_+, y_i \in [0,x_i], y_j \in \mathbb{R}_+, \forall j \neq i \}$
is the set of possible information sets at the beginning of stage three.

Nash equilibrium (NE) of the CPS game is
defined as a strategy profile such that no peer can improve its utility by a unilateral deviation.
The play on the equilibrium path at an NE is called an NE outcome of the CPS game.
A refinement of NE for sequential games
is subgame-perfect equilibrium (SPE), which requires that players choose NE strategies
in any subgame, thereby eliminating incredible threats. Subgame perfection
provides robustness in equilibrium strategies in that
deviation is unprofitable not only at the beginning of the game but also
at any stage of the game. However, formally there is no subgame of
the CPS game starting from stage two or three because
the stage-one choice of a peer is not revealed to other peers. Hence, SPE
fails to provide a refinement of NE in the CPS game.

In order to extend the spirit of subgame perfection to non-singleton information
sets, we can use sequential rationality, which postulates
that players behave optimally in each information set for a given belief system
\cite{fudenberg}. Sequential rationality is required by the solution concepts of perfect Bayesian equilibrium (PBE)
and sequential equilibrium (SE).\footnote{A strategy profile and a belief system
constitute a PBE if the strategies are sequentially rational given
the belief system and the beliefs are updated using Bayes' rule, whenever
possible, given the strategy profile. SE is a refinement of PBE in that
SE perturbs the strategy profile to make Bayes' rule applicable in every
information set. See \cite{fudenberg} for the formal definitions of PBE and SE.}
The difference between these two solution
concepts disappears in the CPS game because the consistent belief of peer $i$ on
$\mathbf{x}_{-i} \triangleq (x_1, \ldots, x_{i-1}, x_{i+1}, \ldots, x_N)$
should be the correct $\mathbf{x}_{-i}$ in both solution concepts.
Hence, we use SE to refer to a solution concept requiring sequential rationality
and specify only the strategy profile to describe SE suppressing the belief system
with an implicit premise that peers hold correct beliefs.\footnote{This requirement can
be relaxed using the notion of self-confirming equilibrium (SCE) \cite{levine}, which requires
only observational consistency in beliefs.}
An SE strategy profile of the CPS game can be found applying a backward induction argument,
which is described in detail in \cite{jpark}. As with NE, the play on the
equilibrium path at an SE is called an SE outcome of the CPS game.

\begin{proposition}
At the unique SE outcome of the CPS game, we have $x_i = \hat{x}_{\kappa}$, $y_i =
0$, $\mathbf{z}_i = (0,\ldots,0)$ for all $i \in \mathcal{N}$. Thus, $w(\mathbf{Z})
= 0$ at SE.
\end{proposition}

\begin{IEEEproof}
A formal proof can be found in \cite[Prop. 1]{jpark}. Since sharing
can incur the cost of upload while it gives no benefit to the sharing peer,
it is never optimal for a peer to share a positive amount. Expecting no sharing, each peer
produces the autarkic optimal amount of content, $\hat{x}_{\kappa}$, which maximizes $f(x) - \kappa x$.
\end{IEEEproof}

\emph{Remark.} Suppose that the cost of production is given by a general function
$c(x_i)$ instead of $\kappa x_i$. There is still no sharing at SE, but each peer produces
an amount that maximizes $f(x) - c(x)$, assuming that a maximum exists.

Proposition 1 shows that when peers behave non-cooperatively,
even a small cost of upload makes the socially valuable P2P
network never utilized because peers are not compensated for their
upload. This result explains the free-riding behavior of peers
in file sharing P2P networks such as Napster and Gnutella,
as reported in \cite{saro}, \cite{adar}.
Using a similar argument as in the formal proof of Proposition 1, we can show that
the NE outcome of the CPS game is the same as the SE outcome. NE may prescribe
suboptimal strategies off the equilibrium path, but there cannot be a positive amount shared
on the equilibrium path.\footnote{A similar remark holds for SCE. As soon as
a peer shares its content, it learns that others request its content, and
thus it will choose not to share at all in order to avoid upload costs.}
Individual utility and total utility at non-cooperative equilibrium are
$f^*(\kappa)$ and $\Pi^{NC} = N f^*(\kappa)$, respectively, where we define
$f^*(\alpha) = \sup_{x \geq 0} \{f(x) - \alpha x\}$
for $\alpha \in \mathbb{R}$ as the conjugate of $f$ \cite{boyd}.\footnote{Note that
the definition of a conjugate is adjusted as $f$ is a concave function.}

\section{Cooperative Schemes}

We consider cooperative schemes in the CPS game, which allow
peers to form collaborative groups and to maximize their joint welfare.
In order to prevent peers from behaving non-cooperatively, cooperative
schemes need to enforce the actions of peers by a contract or a protocol.
The protocol designer can implement a cooperative scheme
if he knows the utility functions of all peers in order to determine a desired operating point
and can enforce the operating point. An example of P2P networks to which a cooperative
scheme can be applied is a camera network, where cameras in different locations capture the images of an
object from various angles.
A property that a desired operating point should possess is Pareto efficiency (PE),
which is satisfied when there is no other operating point that makes some peers
better off without making other peers worse off.
We define social welfare by
the sum of the utilities of peers, i.e., $\Pi(\mathbf{x}, \mathbf{y}, \mathbf{Z}) \triangleq
\sum_{i=1}^N v_i(\mathbf{x}, \mathbf{y}, \mathbf{Z})$. Then an allocation is Pareto
efficient (PE) if it maximizes social welfare among feasible allocations.\footnote{An allocation
$(\mathbf{x}, \mathbf{y}, \mathbf{Z})$ is feasible if $x_i \geq 0, 0 \leq y_i \leq x_i, z_{ii} = 0$,
and $0 \leq z_{ij} \leq y_j$ for all $j \neq i$, for all $i \in \mathcal{N}$.}

\begin{proposition}
Let $\beta \triangleq \frac{1}{N}\kappa + \frac{N-1}{N}(\delta +
\sigma)$. At PE, we have $\sum_{i=1}^N x_i = \hat{x}_{\beta}$ and $x_i = y_i =
z_{ji}$ for all $j \neq i$, for all $i \in \mathcal{N}$. Thus, $w(\mathbf{Z}) =
(N-1) \hat{x}_{\beta}$ at PE.
\end{proposition}

\begin{IEEEproof}
A formal proof can be found in \cite[Prop. 2]{jpark}. PE can occur only
when peers share all produced content and download all shared content.
Then the social welfare maximization problem can be written as
\begin{align*}
\max_{\mathbf{x} \geq 0} N f\left(\sum_{i=1}^N x_i\right) - [\kappa +
(N-1)(\delta + \sigma) ] \sum_{i=1}^N x_i.
\end{align*}
The first-order optimality condition for $X \triangleq \sum_{i=1}^N x_i$
is $f'(X) = \beta$. Note that $\beta$ is the per capita marginal cost of
obtaining one unit of content when $N$ peers share all produced content.
Thus, at PE, the level of total production is chosen to equate the marginal
benefit and the marginal cost of supplying content to every peer in the P2P
network.
\end{IEEEproof}

\emph{Remark.} Proposition 2 determines production up to the aggregate level,
leaving the individual levels unspecified.
This is a by-product of the linear cost function of production
(i.e., constant returns to scale). If we assume a strictly convex cost function
(i.e., decreasing returns to scale), $c(x_i)$, instead of $\kappa x_i$, then PE
requires that every peer produce the same amount $x^o$ that maximizes $f(Nx)
- c(x) - (N-1)(\delta + \sigma)x$, eliminating indeterminacy in the allocation of
total production to peers.

At PE, peers jointly produce $\hat{x}_{\beta}$ and share all produced content so that each peer
consumes the total amount produced. The utility of peer $i$
producing $x_i^o$ at a PE allocation $(\mathbf{x}^o, \mathbf{y}^o, \mathbf{Z}^o)$ is given by
\begin{align}\label{eq:indutil}
v_i(\mathbf{x}^o, \mathbf{y}^o, \mathbf{Z}^o) =
f(\hat{x}_{\beta}) - \delta \hat{x}_{\beta} - [ \kappa + (N-1) \sigma
- \delta] x_i^o.
\end{align}
Note that the utility of a peer is decreasing in its
production level given that the total amount of production is fixed and
that all produced content is shared. Total utility at PE is given by
$\Pi^{PE} = N f^*(\beta)$. Since $\beta < \kappa$ for $N \geq 2$, the consumption of a peer and total utility are
smaller at non-cooperative equilibrium than at PE, i.e., $\hat{x}_{\kappa}
< \hat{x}_{\beta}$ and $\Pi^{NC} < \Pi^{PE}$.

In order to derive a coalitional game \cite{myerson} based on the CPS game,
we need to compute the maximum total utility that a subset of peers can achieve.
Define $\tilde{\beta}(n)$ by
\begin{align*}
\tilde{\beta}(n) = \frac{1}{n}\kappa + \frac{n-1}{n}(\delta + \sigma)
\end{align*}
for $n = 1, 2, \ldots$. Note that $\tilde{\beta}(1) = \kappa$,
$\tilde{\beta}(N) = \beta$, and $\tilde{\beta}(n) \rightarrow \delta + \sigma$ as $n \rightarrow \infty$. $\tilde{\beta}(n)$ can be interpreted as
the per capita marginal cost of obtaining one unit of content when $n$ peers
share all produced content. The maximum total utility
achievable with $n$ peers is given by $G(n) \triangleq n f^*(\tilde{\beta}(n))$, and the
maximum average individual utility achievable with $n$ peers by $g(n) \triangleq G(n)/n = f^*(\tilde{\beta}(n))$. Marginal
product (MP) measures an increment in the maximum total utility when the $n$th peer joins the P2P network, i.e., $MP(1) \triangleq G(1)$ and
$MP(n) \triangleq G(n) - G(n-1) = n f^*(\tilde{\beta}(n)) - (n-1) f^*(\tilde{\beta}(n-1))$
for $n \geq 2$.
The following proposition gives some properties of the functions $g$ and $MP$.

\begin{proposition}
(i) $g(n)$ is increasing in $n$, and $\lim_{n \rightarrow \infty} g(n) = f^*(\delta + \sigma)$.\\
(ii) $MP(n)$ is increasing in $n$, $MP(n) > g(n)$ for all $n \geq 2$, and $\lim_{n \rightarrow \infty} [MP(n) - g(n)] = 0$.
\end{proposition}

Since $g(n)$ is increasing in $n$, there are increasing returns to scale when inputs and outputs are
taken to be peers and total utility, respectively. As there are more peers in the P2P network, the cost
of production can be shared by more peers and the socially valuable P2P network can be utilized
more extensively, which results in an increase in the maximum average individual utility.\footnote{In our model,
there are no congestion effects in that the marginal costs of upload and
download are independent of the number of peers in the network.
If we generalize our model so that the marginal costs of upload and download are increasing in the number of peers,
then additional peers will have not only positive externalities but also negative externalities on the existing peers
and there may exist an optimal network size that maximizes total utility as in \cite{johari}.}
Proposition 3(i) in addition states that the maximum average individual utility is bounded above.
Proposition 3(ii) shows that the no-surplus condition
in the sense of \cite{ostroy} is satisfied only in the limiting case with infinitely many peers. This implies
that the distribution of total utility to peers according to their MP, which is proposed
by the marginal productivity theory of distribution of neoclassical economics, is not feasible
unless there are infinitely many peers. Thus, we rely on cooperative game theory
as an alternative theory of distribution.

Let $\mathcal{S}$ with $\mathcal{S} \neq \emptyset$ and $\mathcal{S}
\subseteq \mathcal{N}$ be a coalition of peers. The characteristic function $v$,
which assigns each coalition the maximum total utility it can create, is given
by
\begin{align}\label{eq:vs}
v(\mathcal{S}) = |\mathcal{S}| f^*(\tilde{\beta}(|\mathcal{S}|)),
\end{align}
where $|\mathcal{S}|$ denotes the number of peers in coalition
$\mathcal{S}$. We set $v(\emptyset) = 0$. A coalitional game is
described by the characteristic function $v$, and we consider
two solution concepts for coalitional games, the core and the Shapley value.
An allocation $(\mathbf{x}, \mathbf{y}, \mathbf{Z})$
has the core property in the coalitional game $v$ if
\begin{align*}
\sum_{i \in \mathcal{N}} v_i(\mathbf{x}, \mathbf{y}, \mathbf{Z}) =
v(\mathcal{N}) \textrm{ and } \sum_{i \in \mathcal{S}}
v_i(\mathbf{x}, \mathbf{y}, \mathbf{Z}) \geq v(\mathcal{S}), \
\forall \ \mathcal{S} \subseteq \mathcal{N}.
\end{align*}
The first condition states that the maximum total utility with the grand coalition $\mathcal{N}$ is distributed to peers (i.e.,
a PE allocation is chosen) while the second condition states that
no coalition can improve the utilities of its members from the current allocation.
Hence, the core describes the stable distributions of total utility
in that no coalition of peers can improve their utilities by separating from the grand coalition.
This implies that, when the protocol designer enforces
an allocation with the core property, no coalition of peers
can object the allocation credibly by threatening to leave the P2P network.
The Shapley value, whose expression can be found in \cite{myerson}, is a distribution of
total utility, $v(\mathcal{N})$, that satisfies a certain set of axioms.
The Shapley value can be considered as a fair distribution of
utility as it takes into account the MP of peers
in all possible orders of arrival in the P2P network.

\begin{proposition}
(i) The core of the coalitional game $v$
is a nonempty convex set whose vertices are given by $(MP(1), MP(2), \ldots, MP(N))$ and all
of its permutations.
At the core, we have $\sum_{i=1}^N x_i = \hat{x}_{\beta}$, $x_i = y_i =
z_{ji}$ for all $j \neq i$, for all $i \in \mathcal{N}$, and
\begin{align}\label{eq:core}
\sum_{i \in \mathcal{S}} x_i \leq
|\mathcal{S}| \frac{f(\hat{x}_{\beta}) - {\delta} \hat{x}_{\beta} - f^*(\tilde{\beta}(|\mathcal{S}|))}{\kappa + (N-1)\sigma -
\delta}
\end{align}
for all $\mathcal{S} \subseteq \mathcal{N}$.\\
(ii) The Shapley value of the coalitional game $v$ is $v_i = f^*(\beta)$ for all $i \in \mathcal{N}$,
which is attained at the symmetric PE allocation,
$x_i = y_i = z_{ji} = \hat{x}_{\beta}/N$ for all $j \neq i$, for all $i \in \mathcal{N}$.
\end{proposition}

\begin{figure}
\centering
\includegraphics[width=0.5\textwidth]{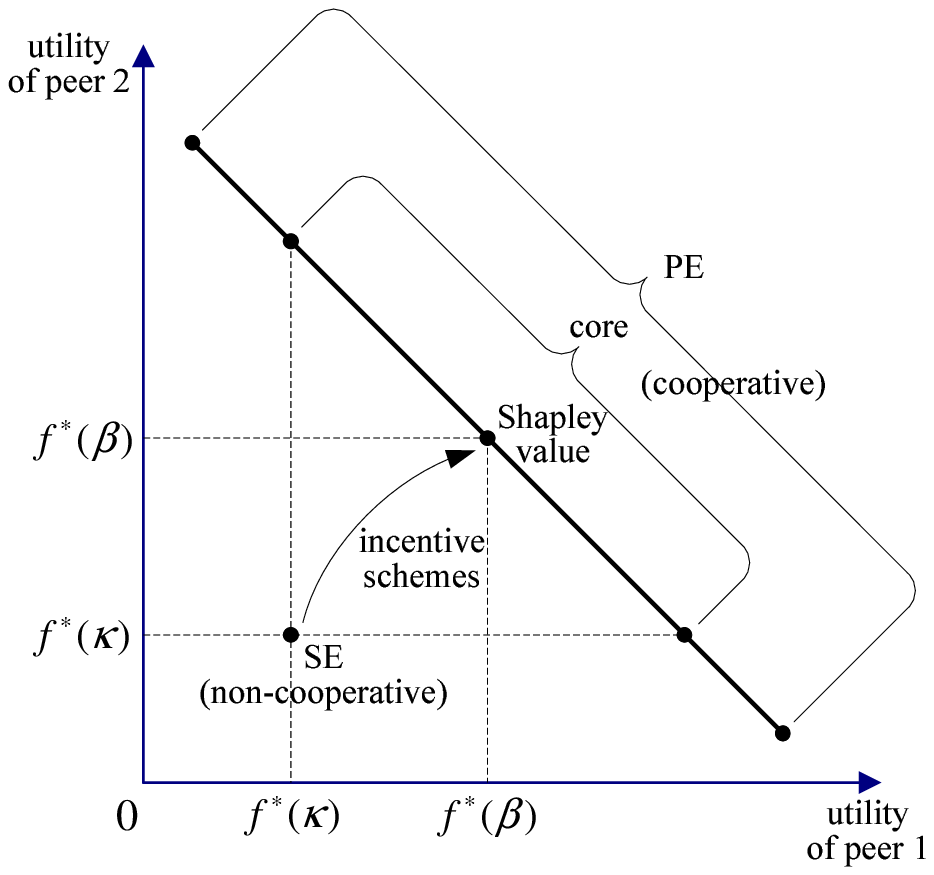}
\caption{Two-peer illustration of utility profiles achieved at non-cooperative and cooperative solution
concepts. Incentive schemes allow non-cooperative peers to achieve a cooperative outcome.}
\label{fig:solcon}
\end{figure}

Fig.~\ref{fig:solcon} illustrates the results in Propositions 1, 2, and 4 with two peers.
Since $\Pi^{NC} < \Pi^{PE}$, PE allocations achieve a higher total utility
than SE allocations. Also, since the core imposes additional constraints on PE,
the core is a subset of PE utility profiles.
Proposition 3(ii) implies that the coalitional game $v$ is convex \cite{shapley},
and thus the results in Proposition 4 can be considered as the corollaries of theorems in \cite{shapley}.
In particular, the core is nonempty and coincides with the unique stable set
in the sense of \cite{von}. Also, the Shapley value is the center of gravity of the core,
which is consistent with the illustration in Fig.~\ref{fig:solcon}. Hence, by prescribing the allocation
that yields the Shapley value, the protocol designer can obtain the stability property
of the core and the fairness property of the Shapley value at the same time.

The maximum utility that a peer can obtain by itself is $f^*(\kappa)$, which can be considered as a
reservation utility. An allocation $(\mathbf{x}, \mathbf{y},
\mathbf{Z})$ satisfies the participation (or individual rationality) constraint for peer
$i$ if $v_i(\mathbf{x}, \mathbf{y}, \mathbf{Z}) \geq f^*(\kappa)$.
An allocation $(\mathbf{x},\mathbf{y}, \mathbf{Z})$ is participation-efficient if it is
PE and satisfies the participation constraint for
every peer. Among PE allocations, condition \eqref{eq:core} for a singleton
coalition $\mathcal{S} = \{i\}$, which can be written as
\begin{align*}
x_i \leq \frac{f(\hat{x}_{\beta}) - {\delta} \hat{x}_{\beta} - f^*(\kappa)}{\kappa + (N-1)\sigma -
\delta},
\end{align*}
is required for the participation constraint for peer $i$.
Since the utility of a peer
decreases in its production level among PE allocations as shown in \eqref{eq:indutil}, the
participation constraint puts an upper bound on the individual
production level to prevent a peer from leaving the P2P network. The core
is a stronger concept than participation-efficiency in that the core
prevents not only a single peer from leaving the P2P network but
also a subset of peers from forming their own P2P network.

Peers choose actions $\mathbf{x}$, $\mathbf{y}$, and $\mathbf{Z}$ over
the three stages of the CPS game.
Suppose that the protocol designer can enforce the sharing levels of
peers in stage two while he cannot enforce
the choices in stages one and three. Then the CPS game is reduced to the CPS game
with enforced sharing levels $\mathbf{y}^e$, where the stage-two choice of peers is
fixed at some $\mathbf{y}^e$.

\begin{proposition}
Suppose that $\hat{x}_{\kappa} \leq \sum_{i=1}^N y_i^e \leq \hat{x}_{\delta}$.
At the SE outcome of the CPS game with enforced sharing levels $\mathbf{y}^e$,
we have  $x_i = y_i^e = z_{ji}$ for all $j \neq i$, for all $i \in \mathcal{N}$.
\end{proposition}

Proposition 5 shows that when peers are required to share $\mathbf{y}^e$
that satisfies $\hat{x}_{\kappa} \leq \sum_{i=1}^N y_i^e \leq \hat{x}_{\delta}$,
they produce exactly the enforced sharing levels and download all shared
content in their self-interest. Since $\hat{x}_{\kappa} \leq \hat{x}_{\beta} \leq \hat{x}_{\delta}$,
the protocol designer can implement a PE allocation by enforcing only the sharing
levels $\mathbf{y}^e$ such that $\sum_{i=1}^N y_i^e = \hat{x}_{\beta}$, leaving
peers to choose the production and download levels on their own.

\section{Payment Schemes}

Suppose that the protocol designer is unable to enforce the sharing levels
of peers. The non-cooperative analysis in Section III suggests that,
in order to avoid the collapse of the P2P network, peers need to be incentivized to
share and upload their produced content.
Pricing is an extensively studied form of incentives to achieve
an efficient use of network resources \cite{varian}.
Pricing schemes have been used in P2P-based web services such as MojoNation in the forms of
tokens and credits.
We say that a pricing scheme is optimal if it induces a PE non-cooperative equilibrium.
In order to determine an optimal pricing scheme,
the protocol designer needs to know the utility functions of peers.
An optimal pricing scheme can be implemented when
the protocol designer can enforce payments and each peer knows the pricing rule applied to it.

In this section, we examine a class of pricing schemes under which the payment to a peer is
increasing in its upload volume and decreasing in its download
volume at the same rate.\footnote{Another class of pricing schemes, called MP pricing schemes,
is proposed and analyzed in \cite{jpark}. Under an MP pricing scheme, payments are determined based on
the sharing levels of peers.} We call such a pricing scheme a linear pricing scheme,
which can be expressed formally as
\begin{align*}
t_i(\mathbf{Z}) = p (u_i - d_i)
\end{align*}
for some price $p > 0$. Note that a linear pricing scheme with any price
satisfies budget balance since it simply transfers payments from downloaders
to uploaders, i.e., $\sum_{i=1}^N t_i(\mathbf{Z}) = p (\sum_{i=1}^N u_i - \sum_{i=1}^N d_i) =
0$ for all $\mathbf{Z}$ and $p$. The
payoff to peer $i$ in the CPS game with the linear pricing scheme with price $p$ is
given by
\begin{align*}
\pi_i(\mathbf{x}, \mathbf{y}, \mathbf{Z}) = v_i(\mathbf{x},
\mathbf{y}, \mathbf{Z}) + t_i(\mathbf{Z}) = f(x_i + d_i) - \kappa x_i - (p + \delta) d_i + (p - \sigma) u_i.
\end{align*}
In effect, the linear pricing scheme with price $p$ increases the
cost of download from $\delta$ to $p + \delta$ and decreases the
cost of upload from $\sigma$ to $\sigma - p$. If the reward for
upload exceeds the cost of upload, i.e., $p > \sigma$, then
peers receive a net benefit from uploading, which provides them with
an incentive for sharing. The following proposition shows that there
exists an optimal pricing scheme in the class of linear pricing
schemes. Moreover, the optimal linear pricing scheme transfers the
utilities of peers so that the equilibrium payoff profile
coincides with the Shapley value for any PE allocation chosen by peers.

\begin{proposition}
Let $p^* = [\kappa + (N-1)\sigma - \delta]/N$. At the SE
outcome of the CPS game with the linear pricing scheme with price
$p^*$, we have $\sum_{i=1}^N x_i = \hat{x}_{\beta}$ and $x_i = y_i = z_{ji}$
for all $j \neq i$, for all $i \in \mathcal{N}$. The payoff of each
peer at SE is given by $f^*(\beta)$.
\end{proposition}

\begin{IEEEproof}
The result follows from Lemma 1 and Proposition 7 of \cite{jpark}.
\end{IEEEproof}

\emph{Remark.} As is the case with PE, indeterminacy in the allocation of
total production to peers can be eliminated by assuming a strictly convex
cost function of production, $c(x_i)$. With a strictly convex
cost function of production, the optimal price $p^*$ can be
found as in Proposition 6 replacing $\kappa$ with $c'(x^o)$, where
$x^o$ is the individual production level at PE as discussed in the remark
following Proposition 2.

A notable feature of the optimal price $p^*$ is that peers are indifferent
between the two alternative methods of obtaining data, production
and download, when they face the optimal price. At PE without
pricing schemes, peers prefer download to production because the marginal
cost of download, $\delta$, is smaller than that of production
and upload, $\kappa+(N-1)\sigma$ (see the coefficient of the term $x_i^o$ in \eqref{eq:indutil}).
The optimal price is chosen such that it equates
the effective marginal cost of download, $p^* + \delta$,
with that of production and upload, $\kappa - (N-1)(p^* -
\sigma)$. As a result, at SE with the optimal linear pricing scheme,
peers obtain the same payoff regardless of their production levels.

With linear cost functions and linear prices,
this is a necessary property of any nondiscriminatory pricing scheme that induces
non-cooperative homogeneous peers to produce a positive bounded amount of content in
aggregate. If the payoff to a peer is increasing in its production level, then
the peer would be willing to produce and upload as much as it can (i.e.,
the peer is overcompensated for its production and the supply of content
is unbounded). On the other hand, if the
payoff to a peer is decreasing in its production level, then the
peer would not produce at
all (i.e., the peer is undercompensated for its production and
the supply of content is zero). The optimal price can be considered as an
equilibrium price of content in that the compensation is
enough to provide incentives for production and upload but
prevents overproduction so that supply equals demand.

Imposing that $y_i = x_i$ and $u_i = (N-1)y_i$ for all $i \in \mathcal{N}$,
the social welfare maximization problem can be written as
\begin{align} \label{eq:primal}
&\max_{\mathbf{x}, \mathbf{d}} \sum_{i=1}^N \left\{f(x_i
+ d_i) - [\kappa + (N-1) \sigma] x_i - \delta d_i \right\} \nonumber \\
&\textrm{subject to } x_i \geq 0, 0 \leq d_i
\leq \sum_{j \neq i} x_j, \textrm{for all $i = 1, \ldots, N$}.
\end{align}
Let $p_i \geq 0$ be a Lagrange multiplier on the constraint $d_i
\leq \sum_{j \neq i} x_j$ for each $i \in \mathcal{N}$, which can
be interpreted as the price that peer $i$ pays for its download.
Then the Lagrangian function can be written as
\begin{align*}
\mathcal{L}(\mathbf{x}, \mathbf{d}; \mathbf{p}) = \sum_{i=1}^N \left\{f(x_i
+ d_i) - [\kappa + (N-1) \sigma] x_i - \delta d_i + p_i(\sum_{j \neq i} x_j - d_i) \right\},
\end{align*}
and the first-order optimality conditions for $x_i$ and $d_i$ are given by
\begin{align*}
f'(x_i + d_i) - \kappa - (N-1) \sigma + \sum_{j \neq i} p_j \leq 0 \textrm{ (with equality if $x_i > 0$)}
\end{align*}
and
\begin{align*}
f'(x_i + d_i) - \delta - p_i \leq 0 \textrm{ (with equality if $d_i > 0$)}.
\end{align*}
At PE, we have $x_i + d_i = \hat{x}_{\beta}$, which yields the Lagrange
multiplier $p_i = p^*$ for all $i \in \mathcal{N}$. The dual decomposition of \eqref{eq:primal} can be written as
\begin{align} \label{eq:seller}
\max_{x_i,d_i \geq 0} f(x_i + d_i) - [\kappa + (N-1) \sigma - \sum_{j \neq i} p_j] x_i - (p_i + \delta) d_i
\end{align}
for each $i$. The solution to \eqref{eq:seller} is given by
\begin{align} \label{eq:dualsol}
&x_i = 0 \textrm{ and } d_i = \hat{x}_{(p_i + \delta)} & \textrm{if $\sum_{j=1}^N p_j < \kappa + (N-1) \sigma - \delta$}&,\nonumber\\
&x_i + d_i = \hat{x}_{(p_i + \delta)} = \hat{x}_{[\kappa + (N-1) \sigma - \sum_{j \neq i} p_j]} & \textrm{if $\sum_{j=1}^N p_j = \kappa + (N-1) \sigma - \delta$}&,\nonumber\\
&x_i = \hat{x}_{[\kappa + (N-1) \sigma - \sum_{j \neq i} p_j]} \textrm{ and } d_i = 0 & \textrm{if $\sum_{j=1}^N p_j > \kappa + (N-1) \sigma - \delta$}&.\footnotemark
\end{align}\footnotetext{We set $\hat{x}_{\alpha} = +\infty$ when $\alpha \leq 0$.}%
Thus, the maximum value of \eqref{eq:seller} is given by $h_i(\mathbf{p}) \triangleq
f^*(\min \{p_i + \delta, \kappa + (N-1) \sigma - \sum_{j \neq i} p_j\})$, and $p_i = p^*$ for all $i$ is the solution
of the dual problem, $\min_{\mathbf{p} \geq 0} \sum_{i=1}^N h_i(\mathbf{p})$.
Thus, a uniform linear pricing scheme suffices to obtain PE allocations.
The problem \eqref{eq:primal} is more general than the resource allocation
problem in \cite{kelly1} in the following aspect. In our problem,
peers can choose to become either a seller or a buyer (or both), and
the amount of resources (i.e., content) supplied in the P2P network is chosen by peers.
On the contrary, in \cite{kelly1}, buyers and sellers are predetermined,
and sellers hold a fixed supply of resources.

If the protocol designer knows the utility
functions of peers, he can compute the optimal price $p^*$ using the expression in Proposition 6.
At the optimal price, equilibrium requires that peers produce $\hat{x}_{\beta}$
in aggregate. Peers can coordinate to achieve total production $\hat{x}_{\beta}$
using the following quantity adjustment process. Initially, each peer $i$ chooses
arbitrary optimal production and download levels $(x_i,d_i)$, which satisfy $x_i + d_i =
\hat{x}_{\beta}$. Peers share their production fully throughout
the process, and thus they can observe the production levels of
other peers indirectly.
If $\sum_{i=1}^N x_i > \hat{x}_{\beta}$
(respectively, $\sum_{i=1}^N x_i < \hat{x}_{\beta}$), then $d_i < \sum_{j \neq i}
x_j$ (respectively, $d_i > \sum_{j \neq i} x_j$) for all $i$. Hence,
if each peer $i$ adjusts its production and download levels by
\begin{align} \label{eq:quantity}
\frac{dx_i}{dt} = -\frac{dd_i}{dt} = \eta_i \left(d_i - \sum_{j \neq i} x_j \right)
\end{align}
for some constant $\eta_i > 0$,\footnote{Since $x_i$ cannot be negative, we assume that peer $i$ stops
adjusting its quantity when $x_i = 0$ and $d_i < \sum_{j \neq i} x_j$.} then the allocation will converge
to an equilibrium allocation, which satisfies $\sum_{i=1}^N x_i = \hat{x}_{\beta}$.

\begin{figure}
\centering
\includegraphics[width=0.6\textwidth]{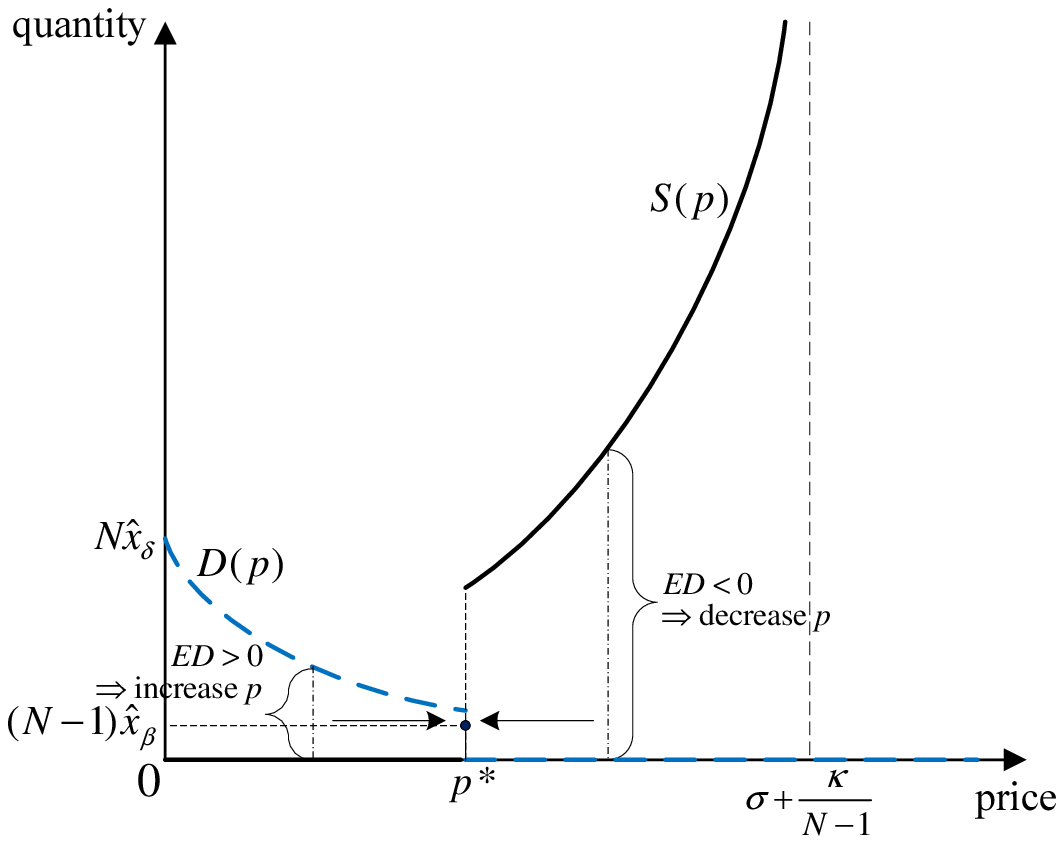}
\caption{Equilibrium interpretation of the optimal price $p^*$. $D(p)$ represents the total
demand for download and $S(p)$ the total supply of upload at price $p$. The equilibrium price
equates demand and supply, i.e., $D(p^*) = S(p^*)$. The protocol designer can reach the equilibrium price by adjusting
the price depending on the excess demand, $ED$.}
\label{fig:linprice}
\end{figure}

Suppose instead that the protocol designer does not know the utility functions of peers.
In this case, the protocol designer can still find the optimal price
by using a price adjustment process similar to that in \cite{kelly2}.
In other words, a price adjustment process can substitute knowledge about
the utility functions of peers.
In the proposed price adjustment process, the protocol designer
announces a price $p$, which applies to every peer.
Given the price, each peer $i$ chooses $(x_i, d_i)$ by solving \eqref{eq:seller}
and reports its choice to the protocol designer. We assume that peers
can coordinate their choices to satisfy $d_i = \sum_{j \neq i} x_j$ for
all $i \in \mathcal{N}$ whenever possible, for example, by using the quantity adjustment
process \eqref{eq:quantity}. The total demand for content at price $p$ is
denoted by $D(p)$ and can be computed as $\sum_{i=1}^N d_i$. Similarly,
the total supply of content at price $p$ is
denoted by $S(p)$ and can be computed as $\sum_{i=1}^N u_i = (N-1) \sum_{i=1}^N x_i$,
as a peer can upload its production up to $(N-1)$ times.
Using \eqref{eq:dualsol}, we obtain
\begin{eqnarray*}
D(p) = \left\{ \begin{array}{ll}
N \hat{x}_{(p + \delta)} & \textrm{if $p < p^*$}\\
(N-1) \hat{x}_{\beta} & \textrm{if $p = p^*$}\\
0 & \textrm{if $p > p^*$}
\end{array} \right.
\textrm{ and}\quad
S(p) = \left\{ \begin{array}{ll}
0 & \textrm{if $p < p^*$}\\
(N-1) \hat{x}_{\beta} & \textrm{if $p = p^*$}\\
N (N-1) \hat{x}_{[\kappa - (N-1) (p-\sigma)]} & \textrm{if $p > p^*$}
\end{array} \right.,
\end{eqnarray*}
as depicted in Fig.~\ref{fig:linprice}.
We define the excess demand at price $p$ by $ED(p) \triangleq D(p) - S(p)$.
The protocol designer adjusts the price following the process
\begin{align*}
\frac{dp}{dt} = \eta ED(p)
\end{align*}
for some constant $\eta > 0$.
Since $ED(p) > 0$ for $p < p^*$ and
$ED(p) < 0$ for $p > p^*$, the price will converge to the optimal
price $p^*$ starting from any initial price.
We compare the above price adjustment process with that in \cite{kelly2}.
In \cite{kelly2}, there are multiple
resources with fixed supply, and each resource manager adjusts
the price of his resource so that aggregate demand for the resource equals
the supply of the resource. In our formulation, by focusing on a uniform linear pricing
scheme, we treat resources provided by different peers as a single
resource. Hence, the protocol designer needs to aggregate demand and supply
by all peers and adjust the price of content
to eliminate excess demand or supply.

We have assumed that peers report their demand and supply truthfully
in the price adjustment process. Suppose instead that peers know
the price adjustment process used by the protocol designer
and can engage in strategic misrepresentation as in \cite{johari2}.
We find that, unlike in \cite{johari2}, no peer can gain from influencing the
equilibrium price by misreporting its demand or supply provided that
other peers report truthfully. In \cite{johari2},
a user, acting as a buyer, can benefit from a lower price of a resource
by underreporting its demand. On the contrary, in our model, a peer is both
a buyer and a seller, and thus it can lower the price only by
increasing its supply, which hurts it as a seller. Peer $i$ can make
the price adjustment process stop at $p' < p^*$ by reporting
$x_i = \hat{x}_{(p'+\delta)}$ and $d_i = 0$.
Since $p'+\delta < \beta < \kappa - (N-1)(p'-\sigma)$, the payoff of peer $i$ is
\begin{align*}
\pi_i = f(\hat{x}_{(p'+\delta)}) - [\kappa-(N-1)(p'-\sigma)]
\hat{x}_{(p'+\delta)} < f(\hat{x}_{(p'+\delta)}) - \beta
\hat{x}_{(p'+\delta)} < f^*(\beta),
\end{align*}
and thus it obtains a lower payoff by manipulating the equilibrium price at $p'$.
For $p > p^*$, the optimal production and download levels for peer $j$
are $x_j = \hat{x}_{[\kappa - (N-1)(p-\sigma)]}$
and $d_j = 0$. Since $d_i \leq \sum_{j \neq i} x_j$, peer $i$ alone cannot
induce $ED(p) = 0$, or $\sum_{i=1}^N d_i = (N-1) \sum_{i=1}^N x_i$, for
some $p > p^*$, if $N > 2$.
When $N = 2$, peer $i$ can make
the price adjustment process stop at $p'' > p^*$ by reporting
$x_i = 0$ and $d_i = \hat{x}_{[\kappa - (N-1)(p''-\sigma)]}$. Since
$\kappa - (N-1)(p''-\sigma) < \beta < p''+\delta$, the payoff of peer
$i$ is
\begin{align*}
\pi_i &= f\left(\hat{x}_{[\kappa - (N-1)(p''-\sigma)]}\right) -
(p''+\delta) \hat{x}_{[\kappa - (N-1)(p''-\sigma)]}\\ &<
f\left(\hat{x}_{[\kappa - (N-1)(p''-\sigma)]}\right) - \beta
\hat{x}_{[\kappa - (N-1)(p''-\sigma)]} < f^*(\beta).
\end{align*}
Again, peer $i$ cannot gain from misreporting.

\section{Differential Service Schemes}

Another form of incentives that encourage sharing by non-cooperative
peers is differential service, in which peers obtain different qualities of
service depending on their contribution levels.
Differential service schemes are widely adopted in
file sharing P2P networks such as BitTorrent \cite{cohen} and KaZaA,
in the forms of tit-for-tat and reputation.
In this section, we capture the differential service in the CPS game
using two modeling approaches based on repeated games and intervention. In a repeated
game model, peers can reciprocate service to each other based on
private or public history. In an intervention model, the system
treats peers differentially based on their contribution to the system.

\subsection{Repeated Game Model}

Suppose that peers interact repeatedly over time in the P2P network. The repeated game
model can support a cooperative outcome among non-cooperative peers
by providing rewards and punishments depending on the past behavior of peers.
The repeated CPS game is a supergame in which the CPS game is played repeatedly.
We use the limit of means criterion \cite{osborne} to evaluate the utility
of a peer in the repeated CPS game to obtain the following result.\footnote{%
A similar result can be obtained with the discounting criterion, in which case
Proposition 7 is restated as ``Any strictly
participation-efficient allocation can be supported as a non-cooperative
equilibrium of the repeated CPS game when peers are sufficiently patient.''}

\begin{proposition}
Any participation-efficient allocation can be supported as a non-cooperative
equilibrium of the repeated CPS game.
\end{proposition}

By Proposition 5, any deviation that is profitable
in the current CPS game involves a deviation in sharing levels, which can be publicly
observed by peers. Hence, the protocol designer can deter
peers from free-riding in the P2P network by making peers play
the SE of the one-shot CPS game in all subsequent CPS games
whenever a peer does not share its required amount of content.
We have assumed that peers serve all the download requests
they receive in stage three. Suppose instead that a peer can choose whether to
upload or not to another peer that requests its content. Then the punishment
following a deviation in sharing levels can be
asymmetric by prescribing peers not to upload to a peer that has ever deviated, which
effectively excludes the deviating peer from the P2P network. Similarly,
refusing a download request from a peer that has not deviated can also be
deterred by using private retaliation (i.e., a non-deviating peer whose request was refused
does the same thing in return to the peer that has refused its request) in all subsequent CPS games.


\subsection{Intervention Model}

Intervention \cite{jpark2} refers to the system
directly influencing the usage of users depending on their behavior.
We consider a particular form of intervention applicable to P2P networks.
Suppose that the P2P network can reduce the download rate of a peer depending
on its rating, where the rating of peer $i$ is defined by its upload to download ratio, i.e., $r_i = u_i/d_i$.
Then a differential service scheme based on intervention
can be described by an intervention function $q: \mathbb{R}_+ \rightarrow \mathbb{R}_+$,
which represents an increase in the marginal cost of download. That is,
when peer $i$ has rating $r_i$, its marginal cost of download after intervention
is given by $\delta + q(r_i)$.
Note that the range of $q$ is constrained to be nonnegative
since we assume that the system can only decrease the download rates.
This imposes a restriction in incentive design compared to a payment scheme,
where it is usually assumed that a payment function can take any positive
or negative real number. However, incentive schemes based on intervention have
advantages in implementation over those based on payment and repeated games.
Unlike a payment scheme, there is no need for transactions in an
intervention scheme since intervention affects peers directly through the system.
Also, intervention can be considered as a substitute of the punishment
strategy in repeated games, but it requires neither repeated interaction among peers
nor the maintenance of history since punishment is executed by the system architecture
rather than by peers.\footnote{For example, the considered type of intervention can be implemented
in a distributed way by
requiring peers to use a certain program to download and upload files, which can adjust
the download rate of a peer automatically based on its past usage.}
We say that an intervention scheme is optimal if it achieves a PE allocation with zero intervention level
at non-cooperative equilibrium. Since any positive level of intervention
results in performance degradation, it is desirable
to have intervention only as a threat, which is called for when misbehavior occurs.

\begin{proposition}
Define an intervention function $q^*$ by $q^*(r_i) = p^*[1-r_i]^+$, where $p^* = [\kappa + (N-1)\sigma - \delta]/N$
and $[r]^+ = \max \{r,0\}$. At the SE outcome of the CPS game with the intervention scheme $q^*$,
we have $x_i = y_i = z_{ji} = \hat{x}_{\beta}/N$ for all $j \neq i$, for all $i \in \mathcal{N}$. Moreover,
$q^*(r_i) = 0$ at SE.
\end{proposition}

\begin{IEEEproof}
As long as $r_i \leq 1$, or $u_i \leq d_i$, for all $i$, the intervention scheme $q^*$ is equivalent to
the optimal linear pricing scheme $p^*$. Since increasing $u_i$ beyond $d_i$ can only increase
the cost of upload without affecting the level of intervention, we must
have $u_i \leq d_i$ for all $i$ at SE. Among SE with the optimal
pricing scheme $p^*$ given in Proposition 6, $u_i \leq d_i$ is satisfied for all $i$ only
when the amount of total production $\hat{x}_{\beta}$ is split equally to all peers.
At this allocation, $u_i = d_i = (N-1)\hat{x}_{\beta}/N$, and thus $r_i = 1$
and $q^*(r_i) = 0$ at SE.
\end{IEEEproof}

Proposition 8 shows that an optimal intervention scheme can be constructed to achieve the
symmetric PE allocation without intervening at equilibrium.
Under the optimal intervention scheme $q^*$, a peer
experiences a reduced download rate whenever it downloads more than it uploads.
Since every peer downloads and uploads the same amount at
the symmetric PE allocation, reduced download rates act only
as a threat at equilibrium, deterring peers from deviation.
The model of \cite{bura} can be considered as using another form of intervention,
where the system determines the proportion of shared content that a peer is allowed to
download as a function of the contribution of the peer.
The model of \cite{vdschaar} can also be interpreted as using an
intervention scheme, where the system no longer serves a peer when its
cumulative average rating falls below a threshold level.
\cite{jpark2} applies an intervention scheme to a multi-user access
network, where the system can jam packets randomly
with a probability that depends on the transmission probabilities
of users. In \cite{jpark2}, intervention affects all users in the system
to the same degree, thus represented by a function that depends
on the actions of all users. On the contrary, intervention considered
in the CPS game, \cite{bura}, and \cite{vdschaar} influences a peer depending only
on its own action, thus allowing the differential service to peers.

\section{Enforced Full Sharing}

We have seen from Proposition 5 that the protocol designer can achieve a PE
allocation by enforcing the sharing levels of peers. As an
alternative scenario, suppose that the protocol designer can enforce
full sharing among peers, but not sharing levels.\footnote{For example, full sharing
can be enforced when there exists an indispensable technology for production and peers have access
to it under the condition of sharing the produced content.} The resulting CPS
game is called the CPS game with enforced full sharing.
Formally, the CPS game with enforced full sharing is a restricted version of
the CPS game where the stage-two choice of each peer $i$ is fixed as $y_i = x_i$.
Note that enforced sharing levels constrain the production decisions of
peers in that peers need to produce at least the required sharing levels.
On the contrary, under enforced full sharing, peers can choose
any levels of production in stage one. The following proposition characterizes
allocations at the SE of the CPS game with enforced full sharing.

\begin{proposition}
Let $\gamma \triangleq \kappa + (N-1)\sigma$. Define $\tilde{x}_{\gamma}$
by $\tilde{x}_{\gamma} = \hat{x}_{\gamma}$ if $\gamma \leq f'(0)$
and $\tilde{x}_{\gamma} =  0$ otherwise (i.e., $\tilde{x}_{\gamma} = \arg \max_{x \geq 0}
\{ f(x) - \gamma x \}$). At the SE outcome of the CPS game with enforced full sharing,
we have $\sum_{i=1}^N x_i = \tilde{x}_{\gamma}$ and $x_i = y_i = z_{ji}$ for
all $j \neq i$, for all $i \in \mathcal{N}$. Thus, $w(\mathbf{Z}) = (N-1)
\tilde{x}_{\gamma}$ at SE with enforced full sharing.
\end{proposition}

\begin{IEEEproof}
A formal proof can be found in \cite[Prop. 3]{jpark}.
Since peers download all shared content at SE,
enforced full sharing increases the effective marginal cost
of production from $\kappa$ to $\kappa + (N-1)\sigma$, which includes
the marginal cost of upload to $(N-1)$ peers.
The stage-one problem for peer $i$ can be written as
\begin{align*}
\max_{x_i \geq 0} f\left(\sum_{i=1}^N x_i\right) - [\kappa + (N-1)\sigma] x_i -
\delta \sum_{j \neq i} x_j
\end{align*}
given $\mathbf{x}_{-i}$, and the result follows.
\end{IEEEproof}

As peers face effectively a higher cost of production with enforced full sharing,
non-cooperative peers reduce their production when full sharing is enforced, i.e., $\tilde{x}_{\gamma} < N \hat{x}_{\kappa}$.
Total utility at SE with enforced full sharing is given by
$\Pi^{FS} = N[f(\tilde{x}_{\gamma}) - {\beta} \tilde{x}_{\gamma}]$.

To make welfare comparisons, we first consider a scenario in which
the number of peers in the P2P network is fixed as $N$.
The price of anarchy (PoA)\footnote{Since the non-cooperative outcome
of the CPS game is unique, the price of anarchy and the price of stability
coincide for the CPS game.} is
defined to be the ratio of social welfare at the worst
non-cooperative equilibrium to that at PE, i.e.,
\begin{align} \label{eq:poa}
PoA \triangleq \frac{\Pi^{NC}}{\Pi^{PE}} = \frac{f^*(\kappa)}{f^*(\beta)}.
\end{align}
The price of no sharing (PoNS) compares social welfare
at SE with and without enforced full sharing, i.e.,
\begin{align*}
PoNS \triangleq \frac{\Pi^{NC}}{\Pi^{FS}} = \frac{f^*(\kappa)}{f(\tilde{x}_{\gamma}) - {\beta} \tilde{x}_{\gamma}}
\textrm{ ($= +\infty$ if $\tilde{x}_{\gamma}=0$)}.
\end{align*}
Finally, the price of underproduction (PoU) compares social
welfare at SE with enforced full sharing and at PE, i.e.,
\begin{align} \label{eq:pou}
PoU \triangleq \frac{\Pi^{FS}}{\Pi^{PE}} = \frac{f(\tilde{x}_{\gamma}) - {\beta}
\tilde{x}_{\gamma}}{f^*(\beta)}.
\end{align}
When $\tilde{x}_{\gamma}>0$, the PoA can be decomposed as the product of
the PoNS and the PoU, i.e., $PoA = PoNS \times PoU$.
The PoA is a widely used measure of the inefficiency of non-cooperative equilibria.
The PoNS measures the welfare implication of enforced full sharing
on selfish peers, and thus it can be used to analyze the value of a technology
that enables enforced full sharing. The PoU measures
inefficiency due to underproduction caused by the selfish behavior of peers
assuming that full sharing is enforced.
The following proposition examines the range of values that each
measure of inefficiency can take when we vary the utility specification of
the model, $f$, $\kappa$, $\delta$, and $\sigma$.

\begin{proposition}
For a fixed size $N \geq 2$ of the P2P network, $PoA \in (0, 1)$, $PoNS \in (0, \infty]$, and $PoU \in [0,1)$. These
bounds are tight.
\end{proposition}

\begin{IEEEproof}
A formal proof can be found in \cite[Prop. 4]{jpark}.
\end{IEEEproof}

Since $\beta < \kappa < \gamma$ for $N \geq 2$, it follows immediately from \eqref{eq:poa}
and \eqref{eq:pou} that $PoA, PoU < 1$, which shows that selfish behavior
results in efficiency losses regardless of whether full sharing is enforced or not.
The relative size of $\Pi^{NC}$ and $\Pi^{FS}$ is
ambiguous, which implies that the enforcement of full sharing may make peers worse
off. This is because enforced full sharing has two offsetting effects
on social welfare. On one hand, full sharing has a positive effect
on welfare by reducing the
cost of obtaining one unit of content to $\beta$, compared
to $\kappa$ in the case of no sharing. On the other hand,
full sharing has a negative effect by increasing the
effective cost of producing one unit of content from $\kappa$ to $\gamma$.
Therefore, the overall welfare implication of enforced full sharing
is determined by the stronger of the two effects.

Next we consider a scenario in which the number of peers in a P2P network
is endogenously determined by peers. There are total $N$ peers that are connected
to each other, and they can form groups to share their content within a group.
The maximum average individual utility increases
with the number of peers in a group as shown in Proposition 3(i). Thus, in a cooperative scenario, peers
will form a P2P network with all the $N$ peers if they accept a new peer as long as
the inclusion of an additional peer
benefits existing peers assuming that peers split total utility equally. In a non-cooperative scenario,
peers do not share content at all, and thus their utilities do
not depend on the number of peers. Hence, the previous results that $\Pi^{PE} = N f^*(\beta)$
and $\Pi^{NC} = N f^*(\kappa)$ are still valid with endogenous network formation. To analyze a
scenario with enforced full sharing, define $\tilde{\gamma}(n)$ by
$\tilde{\gamma}(n) = \kappa + (n-1)\sigma$ for $n = 1,2,\ldots$.
Then the average individual utility of a peer in a P2P network of size $n$
is given by
\begin{align*}
g^{FS}(n) = f(\tilde{x}_{\tilde{\gamma}(n)}) - {\tilde{\beta}(n)}
\tilde{x}_{\tilde{\gamma}(n)}.
\end{align*}
Increasing the size of a P2P network has two opposing effects on average individual
utility. On one hand, increasing the number of peers benefits peers by reducing
the effective marginal cost of obtaining content as represented by $\tilde{\beta}$,
which decreases with $n$. On the other hand, increasing the number of peers
does harm to peers by increasing the effective marginal cost of producing content
as represented by $\tilde{\gamma}$, which increases with $n$. Hence, we can
expect that there exists an optimal size of a P2P network that balances these positive and negative
effects.

Since $g^{FS}(1) = f^*(\kappa) > 0$ and $g^{FS}(n) = 0$ for all
$n \geq (f'(0)-\kappa)/\sigma + 1$, there must exist a maximizer of $g^{FS}(n)$
among $n = 1, \ldots, \lfloor (f'(0)-\kappa)/\sigma + 1 \rfloor$, denoted by $N^*$, where $\lfloor \alpha \rfloor$
is the largest integer smaller than or equal to $\alpha$. We assume that
$N^*$ is unique, which will hold for a generic specification of
the utility function. When $N$ peers form P2P networks endogenously
to maximize their individual utilities, they will
form $\lfloor N/N^* \rfloor$ networks of size $N^*$ and one network of residual peers.
Hence, total utility that a coalition $\mathcal{S}$ can create is given by
\begin{align*}
v^{FS}(\mathcal{S}) = \left\lfloor \frac{|\mathcal{S}|}{N^*} \right\rfloor N^* g^{FS}(N^*) +
\left( |\mathcal{S}| - \left\lfloor \frac{|\mathcal{S}|}{N^*} \right\rfloor \right)
g^{FS}\left( |\mathcal{S}| - \left\lfloor \frac{|\mathcal{S}|}{N^*} \right\rfloor \right).
\end{align*}
In order to examine the stability property of endogenous network formation, we
characterize the core of the coalitional game $v^{FS}$.

\begin{proposition}
Suppose that $N^* < N$.
If $N$ is a multiple of $N^*$, then the core of the coalitional game $v^{FS}$ consists of
a unique element $v_i = g^{FS}(N^*)$ for all $i \in \mathcal{N}$. Otherwise, the
core is empty.
\end{proposition}

Note that we necessarily have $N^* < N$ when $N \geq (f'(0)-\kappa)/\sigma + 1$.
When $N$ is not a multiple of $N^*$, there is a residual network, whose
size is smaller than $N^*$. A peer in the residual network can bid a utility
smaller than $g^{FS}(N^*)$ to form a network of size $N^*$ including itself,
yielding instability for the networks of size $N^*$.
Suppose that $N$ is a multiple of $N^*$ so that
the core is nonempty. The utility profile in the core is achieved by peers forming $N/N^*$
networks, producing $x_i = \hat{x}_{\tilde{\gamma}(N^*)}/N^*$ for all $i$,
and sharing all produced content within a network. Social
welfare at the allocation with the core property is
$\Pi^{FS} = N g^{FS}(N^*)$. Since $g^{FS}(N^*) \geq g^{FS}(1) = f^*(\kappa)$,
we have $PoNS \leq 1$ when peers can form P2P networks of the optimal size.
That is, with endogenous network formation, enforced full sharing can only improve
the welfare of peers because peers are given the option of operating in an autarkic
manner.

\section{Numerical Illustration}

\begin{figure}%
\centering
\subfloat[][]{%
\label{fig:num-a}%
\includegraphics[width=0.49\textwidth]{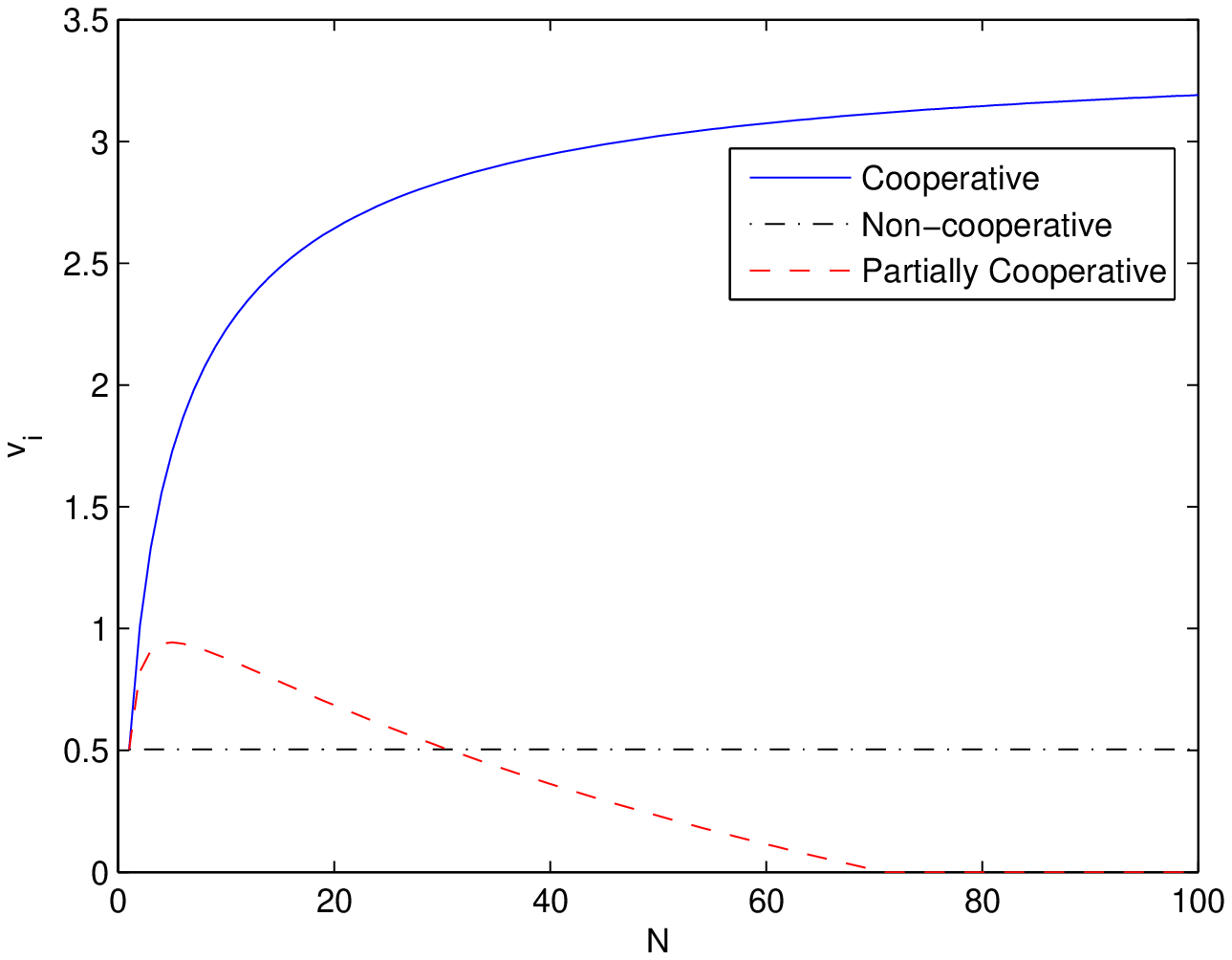}}%
\hspace{8pt}%
\subfloat[][]{%
\label{fig:num-b}%
\includegraphics[width=0.49\textwidth]{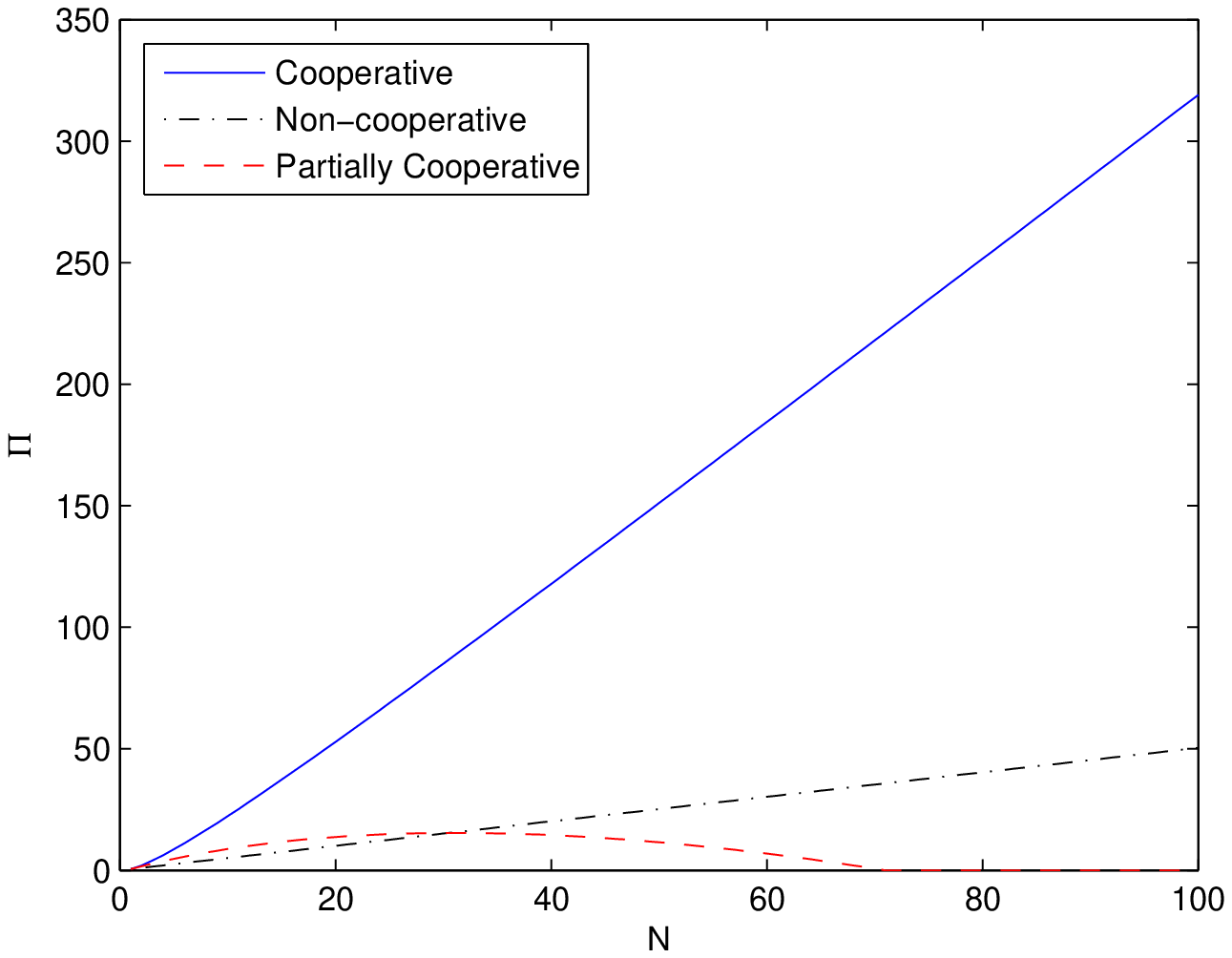}}\\
\subfloat[][]{%
\label{fig:num-c}%
\includegraphics[width=0.49\textwidth]{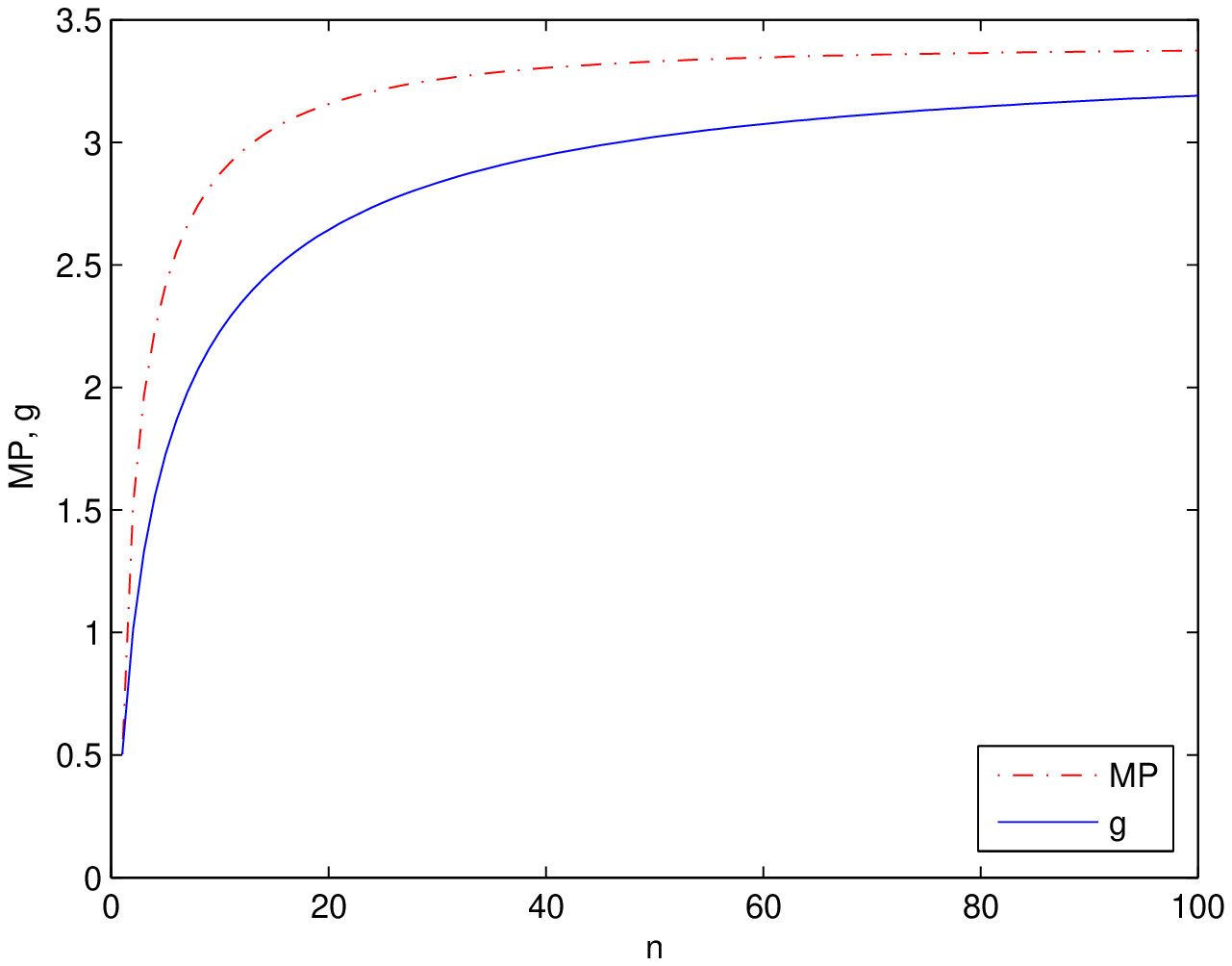}}%
\hspace{8pt}%
\subfloat[][]{%
\label{fig:num-d}%
\includegraphics[width=0.49\textwidth]{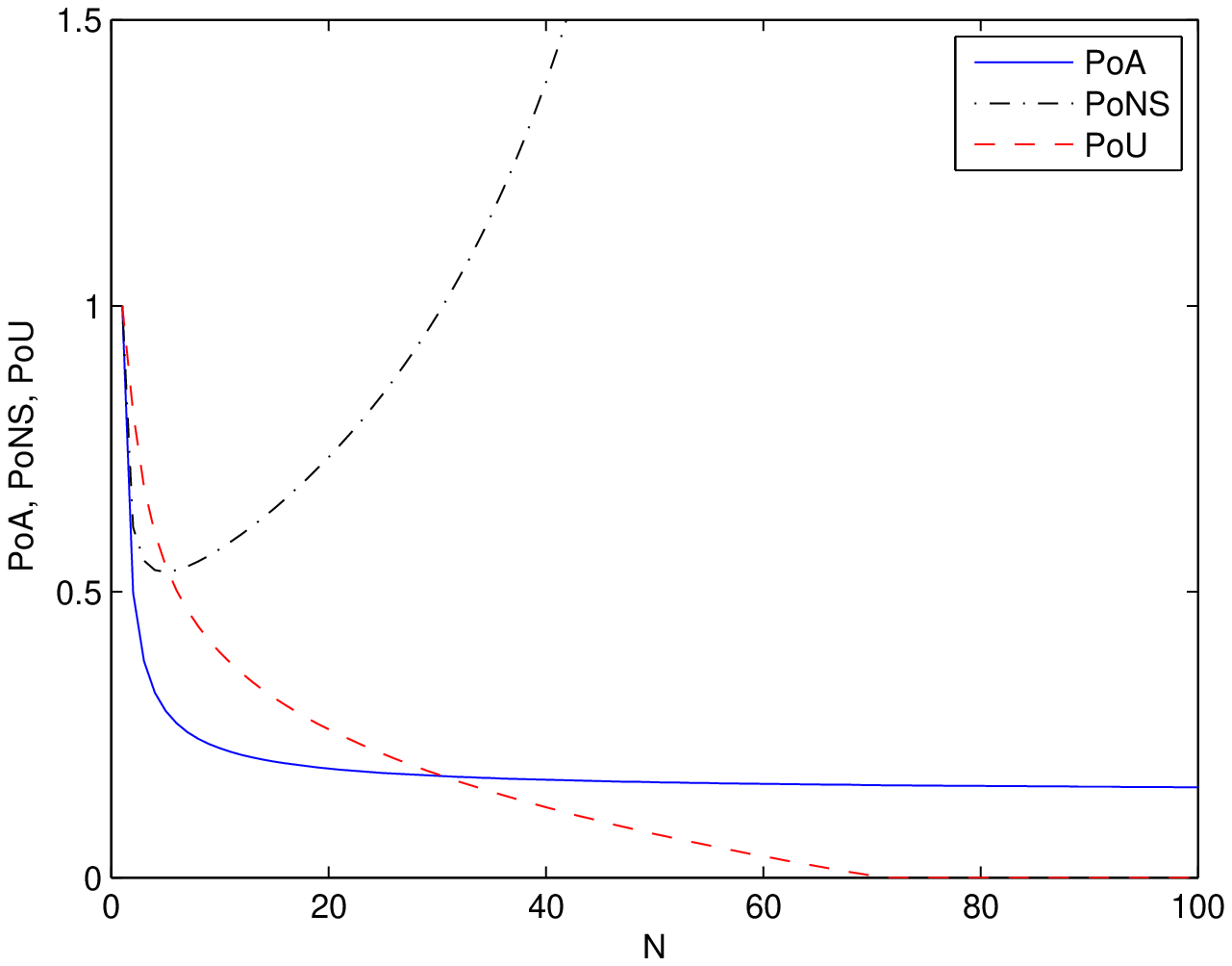}}\\
\subfloat[][]{%
\label{fig:num-e}%
\includegraphics[width=0.49\textwidth]{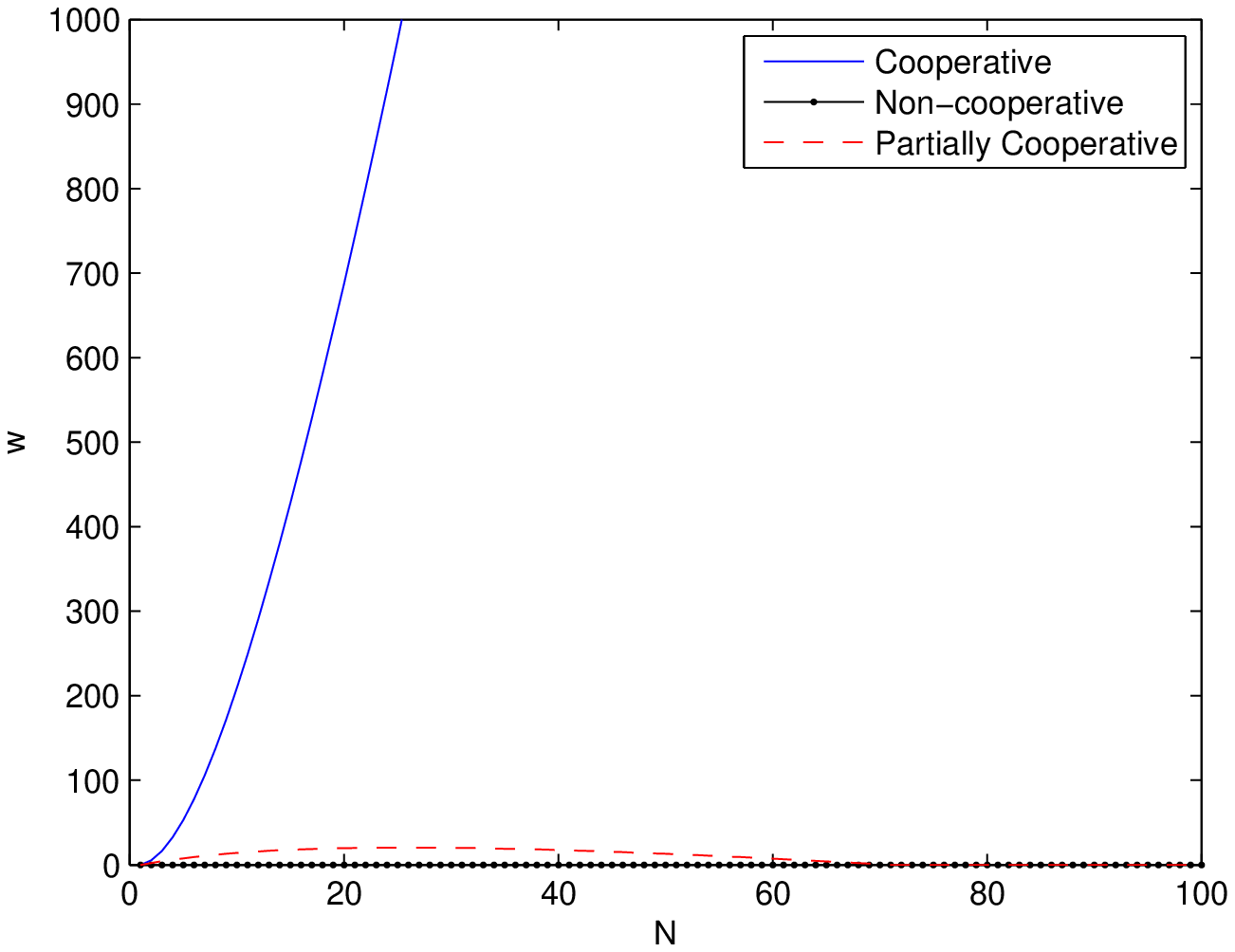}}%
\hspace{8pt}%
\subfloat[][]{%
\label{fig:num-f}%
\includegraphics[width=0.49\textwidth]{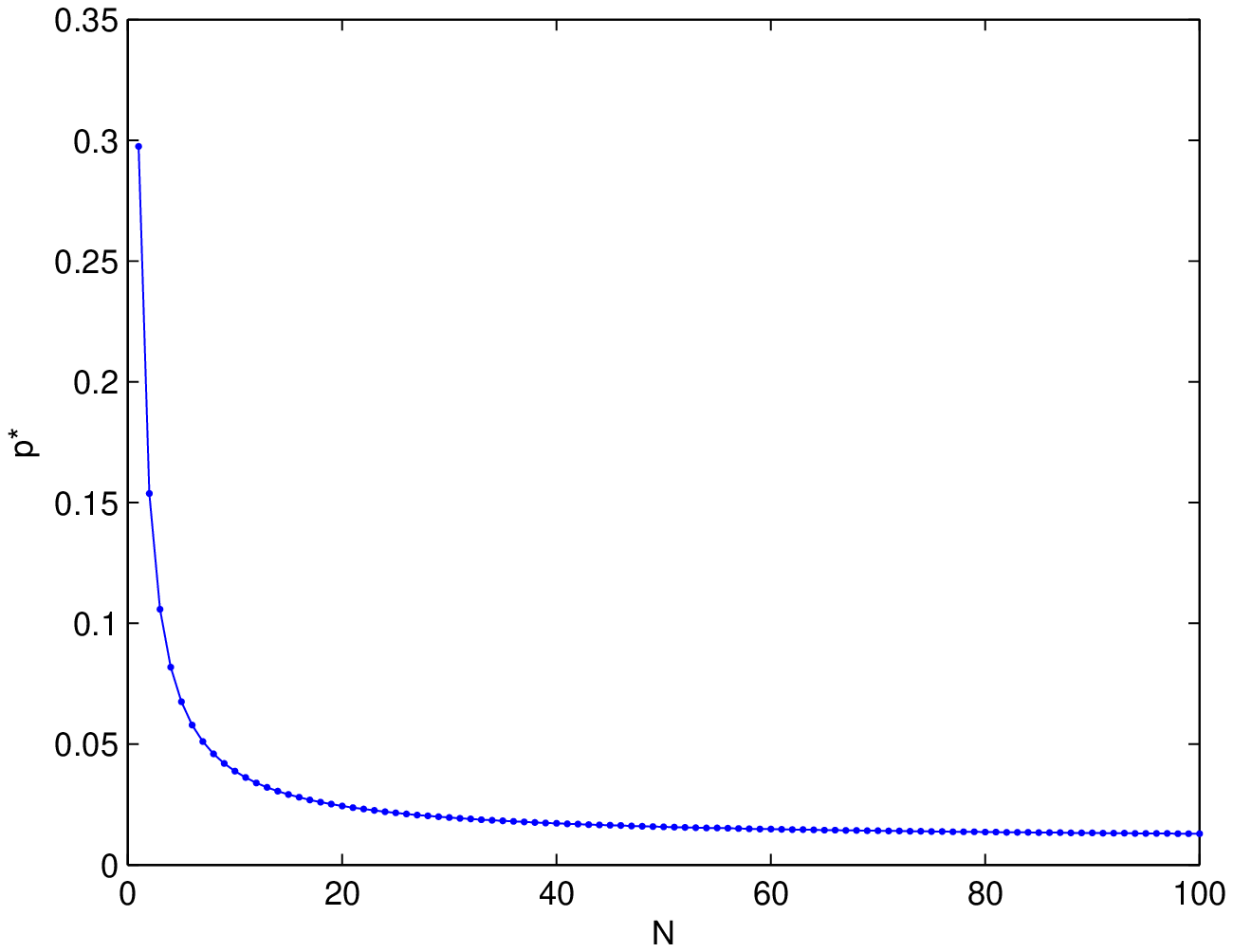}}%
\vspace{-8pt}%
\caption[Numerical illustration.]{Numerical results with $f(x) = log (1+x)$, $\kappa = 0.3$, $\delta =
0.0025$, $\sigma = 0.01$, and varying $N$ from 1 to 100:
\protect\subref{fig:num-a} average individual utility,
\protect\subref{fig:num-b} total utility,
\protect\subref{fig:num-c} marginal product and the maximum average individual utility,
\protect\subref{fig:num-d} inefficiency measures,
\protect\subref{fig:num-e} utilization of the P2P network, and
\protect\subref{fig:num-f} optimal linear prices.
}%
\label{fig:num}%
\end{figure}

In this section, we provide illustrative results using a particular
utility specification and varying the number of peers. For the utility
function of peers, we use $f(x) = log (1+x)$, $\kappa = 0.3$, $\delta =
0.0025$, and $\sigma = 0.01$.\footnote{The authors of \cite{goyal} use the same benefit function
for their illustrative examples.}
We consider the (exogenous) number of peers in the P2P
network, $N$, from 1 to 100. Fig.~\ref{fig:num}\subref{fig:num-a} shows
average individual utility in the three scenarios: $f^*(\beta)$ in the
cooperative case, $f^*(\kappa)$ in the non-cooperative case, and
$f(\tilde{x}_{\gamma}) - \beta \tilde{x}_{\gamma}$ in the partially cooperative case (i.e., enforced full
sharing). It can be seen that $f^*(\beta)$ is increasing in $N$,
verifying Proposition 3(i), that $f^*(\kappa)$ is independent of $N$,
and that $f(\tilde{x}_{\gamma}) - \beta \tilde{x}_{\gamma}$ reaches a peak
at $N = 5$ and is zero for all $N \geq 71$. Fig.~\ref{fig:num}\subref{fig:num-b}
plots total utility in the three scenarios: $\Pi^{PE}$ in the
cooperative case, $\Pi^{NC}$ in the non-cooperative case, and
$\Pi^{FS}$ in the partially cooperative case.

Fig.~\ref{fig:num}\subref{fig:num-c} compares the MP of the $n$th peer, $MP(n)$,
with the maximum average individual utility that $n$ peers can achieve, $g(n)$, verifying
Proposition 3(ii). Fig.~\ref{fig:num}\subref{fig:num-d} plots the three inefficiency
measures defined in Section VII. Since $f^*(\kappa)$ is independent of $N$, we
can see that the PoA and the PoNS change with $N$ in the opposite way that
$f^*(\beta)$ and $f(\tilde{x}_{\gamma}) - \beta \tilde{x}_{\gamma}$ change, respectively.
Since $f^*(\beta)$ converges to $f^*(\delta + \sigma) = 3.3945$, the PoA converges to
$f^*(\kappa)/f^*(\delta + \sigma) = 0.1485$ as $N$ goes to infinity. Fig.~\ref{fig:num}\subref{fig:num-e}
shows the utilization of the P2P network in the three scenarios: $(N-1) \hat{x}_{\beta}$ in the
cooperative case, $0$ in the non-cooperative case, and
$(N-1) \hat{x}_{\gamma}$ in the partially cooperative case.
We can see no utilization in the non-cooperative case and underutilization
(and no utilization for $N \geq 71$) in the partially cooperative case
compared to the cooperative case, which exhibits a high utilization of the
P2P network. Finally, Fig.~\ref{fig:num}\subref{fig:num-f} plots the optimal
linear price $p^*$ as a function of $N$. As can be seen its expression in
Proposition 6, $p^*$ decreases with $N$ and converges to $\sigma = 0.01$
as $N$ goes to infinity.

\section{Conclusion and Future Directions}

In this paper, we have provided a unified framework to investigate
incentive issues in content production and sharing using various game
theoretic approaches. We have characterized the non-cooperative and cooperative outcomes of
the CPS game and have shown that incentive schemes such as payment schemes
and differential service schemes can yield a cooperative outcome
among non-cooperative peers. Throughout the paper, we have
discussed enforcement and information requirements
to implement the solutions of different approaches with a protocol. Our analysis
allows protocol designers to compare the performance and the overheads\footnote{In this paper,
we have mainly focused on communication and informational overheads, and have not addressed
the issues of the complexity of computing different
game theoretic solutions. Some complexity issues can be found in \cite{nisan}.}
of different approaches and eventually helps them select the best approach
given a network environment they face.

We have maintained the homogeneity assumption in order to keep our model tractable
so that we can better illustrate different approaches by providing
analytic results. However, all the concepts in this paper can be straightforwardly
applied to the case of peers with heterogeneous utility functions.
The convexity of the coalitional game in the cooperative approach will still hold, but computing
cooperative solutions will become more complicated with heterogeneous peers.
Non-cooperative solutions will remain the same
with a minor change that individually optimal
production levels will differ across peers. Also, heterogeneous peers
in a distributed system offer a natural scenario to which a mechanism
design approach can be applied. Lastly, the protocol
designer may need to discriminate heterogeneous peers in order to achieve a cooperative
outcome using a linear pricing scheme. Investigating how the results in this paper extend to
and change in P2P networks with heterogeneous peers will provide interesting and challenging
future research directions.

\appendices

\section{Benefit Function Proportional to the Amount of Distinct Files}

For simplicity, suppose that peers choose the number of files from the set of
nonnegative integers. There are total $M$ files that can be potentially produced by a peer,
where $M$ is a large positive integer. When
peer $i$ produces $x_i$ files, it draws $x_i$ files with
replacement from the $M$ files. (Now $\kappa$ can be considered as a constant cost of a draw.)
Each file is drawn with equal probability of $1/M$. Let $c_i$ be the number of files that peer $i$
consumes, i.e., $c_i = x_i + d_i$. Since peers cannot identify the
content of files produced by others before download, $c_i$ files that peer $i$ consumes can be considered as $c_i$
independent draws from the $M$ files. The probability that a given
file is not one of the $c_i$ files is $\left(1 - \frac{1}{M} \right)^{c_i}$.
Hence, the expected number of distinct files in the $c_i$ files is
\begin{align*}
M \left[ 1 - \left(1 - \frac{1}{M} \right)^{c_i} \right].
\end{align*}
Since $\left(1 - \frac{1}{M} \right)^{c_i} \geq 1 - \frac{c_i}{M}$,
we have
\begin{align*}
M \left[ 1 - \left(1 - \frac{1}{M} \right)^{c_i} \right] \leq c_i,
\end{align*}
and $c_i - M \left[ 1 - \left(1 - \frac{1}{M} \right)^{c_i} \right]$
is the expected number of redundant files in the $c_i$ files. If the
benefit of consumption is proportional to the
number of distinct files, the expected benefit function is given by
\begin{align*}
f(c_i) = a M \left[ 1 - \left(1 - \frac{1}{M} \right)^{c_i} \right]
\end{align*}
for some constant $a > 0$. Note that $f$ satisfies all the assumptions for a benefit
function given in Section II when $a > \kappa$ if we take $c_i$ as a nonnegative real number
rather than a nonnegative integer. In particular, $f(0) = 0$ and $f$ is
concave.

\section{Proofs of Propositions}

\noindent \textbf{Proof of Proposition 3.} (i) Note that $g(n)
= f^*(\tilde{\beta}(n)) = f^* \circ \tilde{\beta} (n)$. Also, $f^*(\alpha)
= f(\hat{x}_{\alpha}) - \alpha \hat{x}_{\alpha}$, where $f'(\hat{x}_{\alpha}) = \alpha$,
for $\alpha \in (0,f'(0)]$. That is, $\hat{x}_{\alpha}$ is the unique maximizer
of $f(x) - \alpha x$ on $\mathbb{R}_+$. Choose $\alpha_1, \alpha_2 \in (0,f'(0)]$ such that $\alpha_1 < \alpha_2$.
Then $f^*(\alpha_2) = f(\hat{x}_{\alpha_2}) - \alpha_2 \hat{x}_{\alpha_2}
< f(\hat{x}_{\alpha_2}) - \alpha_1 \hat{x}_{\alpha_2} < f(\hat{x}_{\alpha_1}) - \alpha_1 \hat{x}_{\alpha_1}
= f^*(\alpha_1)$. Hence, $f^*$ is decreasing on $(0,f'(0)]$. Since
\begin{align*}
\tilde{\beta}(n) = \frac{1}{n} (\kappa - \delta - \sigma) + \delta + \sigma,
\end{align*}
$\tilde{\beta}$ is decreasing in $n$ and its range lies in $(\delta+ \sigma, \kappa]
\subset (0,f'(0)]$. Since $g$ is a composite function of two decreasing functions,
it is increasing.

Since $f$ is closed and strictly concave on $\mathbb{R}_{+}$, $f^*$ is differentiable on $\mathbb{R}_{++}$ \cite{boyd}.
Then $f^*$ is continuous on $\mathbb{R}_{++}$, and thus $\lim_{n \rightarrow \infty} g(n) = \lim_{n \rightarrow \infty} f^*(\tilde{\beta}(n))
= f^*(\lim_{n \rightarrow \infty} \tilde{\beta}(n)) = f^*(\delta + \sigma)$.

(ii) To prove that $MP$ is increasing in $n$, it suffices to show the strict convexity of $G$,
taking the domain of $\tilde{\beta}$ and $G$ as $\mathbb{R}_{++}$ instead of $\{1,2,\ldots\}$.
Since $G(n) = n f^*(\tilde{\beta}(n))$ and $f^*$ and $\tilde{\beta}$ are differentiable on $\mathbb{R}_{++}$,
by the chain rule $G$ is differentiable and
\begin{align*}
G'(n) = f^*(\tilde{\beta}(n)) + n (f^*)'(\tilde{\beta}(n)) \tilde{\beta}'(n).
\end{align*}
Note that $(f^*)' = -(f')^{-1}$
on $(0,f'(0))$, $f'$ is continuously differentiable on $\mathbb{R}_{++}$,
and $f''(x) \neq 0$ for all $x \in \mathbb{R}_{++}$. By the inverse function theorem,
$f^*$ is twice continuously differentiable on $(0,f'(0))$, and we have
\begin{align*}
G''(n) = (f^*)'(\tilde{\beta}(n)) [ 2 \tilde{\beta}'(n) + n \tilde{\beta}''(n)] + n (f^*)''(\tilde{\beta}(n)) (\tilde{\beta}'(n))^2.
\end{align*}
Since $2 \tilde{\beta}'(n) + n \tilde{\beta}''(n) = 0$ and $f^*$ is strictly convex on $(0,f'(0))$, we have
$G''(n) = n (f^*)''(\tilde{\beta}(n)) (\tilde{\beta}'(n))^2 > 0$ for all $n \in \mathbb{R}_{++}$. Thus,
$MP(n)$ is increasing in $n$.

Note that $MP(n) - g(n) = (n-1) [g(n) - g(n-1)]$ for $n \geq 2$.
Since $g$ is increasing, we have $MP(n) - g(n) > 0$ for all $n \geq 2$.

Since $f^*$ is convex and differentiable, we have
\begin{align*}
f^*(\tilde{\beta}(n)) - f^*(\tilde{\beta}(n-1)) &\leq -(f^*)'(\tilde{\beta}(n)) [\tilde{\beta}(n-1) - \tilde{\beta}(n)]\\
&= \hat{x}_{\tilde{\beta}(n)} [\tilde{\beta}(n-1) - \tilde{\beta}(n)]\\
&< \hat{x}_{(\delta + \sigma)} [\tilde{\beta}(n-1) - \tilde{\beta}(n)].
\end{align*}
Note that
\begin{align*}
(n-1) [\tilde{\beta}(n-1) - \tilde{\beta}(n)] = \frac{1}{n} (\kappa - \delta - \sigma).
\end{align*}
Hence,
\begin{align*}
0 < (n-1) [g(n) - g(n-1)] < \hat{x}_{(\delta + \sigma)} \frac{1}{n} (\kappa - \delta - \sigma),
\end{align*}
and taking limits as $n \rightarrow \infty$ yields the desired result.\hfill$\blacksquare$

\bigskip
\noindent \textbf{Proof of Proposition 4.} (i) Proposition 3(ii)
implies that the coalitional game $v$ is convex. Hence,
the first sentence follows from theorems 3 and 4 of \cite{shapley}.
The first condition for the core property, $\sum_{i \in \mathcal{N}}
v_i(\mathbf{x}, \mathbf{y}, \mathbf{Z}) = v(\mathcal{N})$, is
equivalent to PE, which requires $\sum_{i=1}^N x_i =
\hat{x}_{\beta}$, $x_i = y_i = z_{ji}$ for all $j \neq i$, for all
$i \in \mathcal{N}$ as shown in Proposition 2.
Choose an arbitrary coalition $\mathcal{S}$. For a PE
allocation $(\mathbf{x}, \mathbf{y}, \mathbf{Z})$, we have
\begin{align*}
\sum_{i \in \mathcal{S}} v_i(\mathbf{x}, \mathbf{y}, \mathbf{Z}) =
f(\hat{x}_{\beta}) - \delta \hat{x}_{\beta} - [ \kappa + (N-1) \sigma -
\delta] \sum_{i \in \mathcal{S}} x_i.
\end{align*}
Hence, using \eqref{eq:vs}, we can show that \eqref{eq:core} is
equivalent to the second condition for the core property, $\sum_{i \in
\mathcal{S}} v_i(\mathbf{x}, \mathbf{y}, \mathbf{Z}) \geq
v(\mathcal{S})$.

(ii) Let $(v_1, \ldots, v_N)$ be the Shapley value of the coalitional game $v$.
By the efficiency property of
the Shapley value, we have $\sum_{i=1}^N v_i = v(\mathcal{N})$.
Also, by the symmetry axiom, we have $v_i = v_j$ for all $i, j \in \mathcal{N}$.
Combining these two yields $v_i = v(\mathcal{N})/N = f^*(\beta)$ for all $i \in \mathcal{N}$. Using \eqref{eq:indutil}, we can see
that $x_i = \hat{x}_{\beta}/N$ is necessary to obtain the Shapley value.\hfill$\blacksquare$

\bigskip
\noindent \textbf{Proof of Proposition 5.} Suppose that $d_i < \sum_{j \neq i}
y_j^e$ for some $i \in \mathcal{N}$ at SE. Then it must be the case that $x_i + \sum_{j \neq i}
y_j^e > \hat{x}_{\delta}$. Since $\sum_{j \neq i} y_j^e \leq \hat{x}_{\delta}$, we have
$x_i > 0$. Then we obtain a contradiction to SE because peer $i$ can improve
its utility by reducing $x_i$ and increasing $d_i$ by the same amount. Thus, $z_{ij} = y_j^e$
for all $j \neq i$, for all $i \in \mathcal{N}$.

The requirement for peer $i$ that $y_i = y_i^e$ in stage two restricts its stage-one choice
with $x_i \geq y_i^e$. Suppose that $x_i > y_i^e$ for some $i \in \mathcal{N}$ at SE.
Since $d_i = \sum_{j \neq i} y_j^e$ at SE, the first-order effect of increasing $x_i$
on $v_i$ is given by $\partial v_i / \partial x_i = f'(x_i + \sum_{j \neq i} y_j^e) - \kappa$.
Since $x_i > y_i^e$ implies $x_i + \sum_{j \neq i} y_j^e > \hat{x}_{\kappa}$, we have
$\partial v_i / \partial x_i < 0$ for $x_i > y_i^e$, and thus peer $i$ becomes worse off by choosing $x_i > y_i^e$,
contradicting SE.\hfill$\blacksquare$

\bigskip
\noindent \textbf{Proof of Proposition 7.} Let $(\mathbf{x}^o, \mathbf{y}^o, \mathbf{Z}^o)$
be a participation-efficient allocation.
Consider the following repeated game strategy for peer $i$: start with a cooperative strategy
in the CPS game $x_i = x_i^o$, $y_i(x_i) = x_i$,
and $\mathbf{z}_i(x_i,\mathbf{y}) = \mathbf{z}_i^*(x_i,\mathbf{y})$, where
$\mathbf{z}_i^*(x_i,\mathbf{y})$ is the optimal download profile of peer $i$ given $(x_i,\mathbf{y})$,
play the cooperative strategy if
$\mathbf{y} = \mathbf{y}^o$ in all the previous CPS games, and play
the SE strategy of the one-shot CPS game, i.e., $x_i = \hat{x}_{\kappa}$, $y_i(x_i) = 0$,
and $\mathbf{z}_i(x_i,\mathbf{y}) = \mathbf{z}_i^*(x_i,\mathbf{y})$, if
$\mathbf{y} \neq \mathbf{y}^o$ in at least one of the previous CPS games.
Proposition 5 implies that peer $i$ cannot gain in the current
CPS game by deviating to $x_i > x_i^o$ or $z_{ij} < y_j^o$ for some $j \neq i$.
Hence, a profitable deviation involves either $x_i < x_i^o$ in stage one or $y_i < y_i^o$ in stage two (or both).
Either case results in a reduction in the sharing level from $y_i^o$.
Since sharing levels are publicly observed, any profitable deviation
is detectable and punishment will be triggered. Since the gain from
deviation in the current CPS game is bounded above, it will be erased by the
punishment in the long run. In other words, peer $i$ receives $v_i^o = v_i(\mathbf{x}^o, \mathbf{y}^o, \mathbf{Z}^o)$
on average if it follows the described repeated game strategy and $f^*(\kappa)$ if it deviates in a way
that the deviation increases the current utility. Since $(\mathbf{x}^o, \mathbf{y}^o, \mathbf{Z}^o)$
is participation-efficient, we have $v_i^o \geq f^*(\kappa)$ for all $i \in \mathcal{N}$.
Hence, the described repeated game strategy, which realizes the allocation
$(\mathbf{x}^o, \mathbf{y}^o, \mathbf{Z}^o)$ in every CPS game, is a non-cooperative equilibrium of
the repeated CPS game.\hfill$\blacksquare$

\bigskip
\noindent \textbf{Proof of Proposition 11.}
Assume that the core is nonempty and choose a utility profile $\mathbf{v} = (v_1,\ldots,v_N)$ in the core. Suppose that there exists a peer $i$ with $v_i <
g^{FS}(N^*)$. Consider a coalition $\mathcal{S}$ of size $N^*$ that do
not include peer $i$, which must exist since $N > N^*$. Then
$\sum_{j \in \mathcal{S}} v_j = N^* g^{FS}(N^*)$, and thus
$\sum_{j \in \mathcal{S} \setminus \{k\}} v_j \leq (N^*-1) g^{FS}(N^*)$, where
peer $k$ is the one that receives the highest utility among peers in $\mathcal{S}$.
Then $\sum_{j \in (\{i\} \bigcup \mathcal{S} \setminus \{k\})} v_j < N^* g^{FS}(N^*)$,
and thus peer $i$ and peers in $\mathcal{S} \setminus \{k\}$ can block the utility profile $\mathbf{v}$.
Hence, we need to have $v_i \geq g^{FS}(N^*)$ for all $i \in \mathcal{N}$. This is possible, with equality, only if
$N$ is a multiple of $N^*$. We can confirm that the core is nonempty since
the utility profile $v_i = g^{FS}(N^*)$ for all $i \in \mathcal{N}$ satisfies
the definition of the core.\hfill$\blacksquare$

\end{document}